\documentclass[twocolumn,preprintnumbers, superscriptaddress, preprint
longbibliography, prr]{revtex4-1}
\usepackage{amsmath}
\usepackage{stmaryrd}
\usepackage{txfonts}
\usepackage{amssymb}
\usepackage{mathrsfs}
\usepackage{graphicx}
\usepackage{dcolumn}

\usepackage{bm}
\usepackage{braket}
\usepackage{epsfig}
\usepackage{color}
\usepackage{ulem}
\usepackage{bibunits}

\usepackage[colorlinks=true, linkcolor=blue, citecolor=blue, urlcolor=blue]{hyperref}
\newcommand{\SZL}[1] {{\color{black}#1}}

\begin{document}

\title{Enhanced Superconductivity in quasi-periodic crystals}

\author{Zhijie Fan}
\affiliation{Theoretical Division, T-4 and CNLS, Los Alamos National Laboratory, Los Alamos, New Mexico 87545, USA}
\affiliation{Department of Physics, University of Virginia, Charlottesville, VA 22904, USA}

\author{Gia-Wei Chern}
\affiliation{Department of Physics, University of Virginia, Charlottesville, VA 22904, USA}

\author{Shi-Zeng Lin}
\affiliation{Theoretical Division, T-4 and CNLS, Los Alamos National Laboratory, Los Alamos, New Mexico 87545, USA}

\date{\today}

\begin{abstract}
We study superconductivity in a family of one dimensional incommensurate system with $s$-wave pairing interaction. The incommensurate potential can alter the spatial characteristics of electrons in the normal state, leading to either extended, critical, or localized wave functions. We find that superconductivity is significantly enhanced when the electronic wave function exhibits a critical multifractal structure. This criticality also manifests itself in the power-law dependence of superconducting temperature on the pairing strength. As a consequence, an extended superconducting domain is expected to exist around the localization-delocalization transition, which can be induced by either tuning the amplitude of the incommensurate potential, or by varying the chemical potential across a mobility edge. Our results thus suggest a novel approach to enhance superconducting transition temperature through engineering of incommensurate potential.  
\end{abstract}

\maketitle

\section{Introduction}
Electronically incommensurate potential appears in many condensed matter systems. Prominent examples include quasicrystals, borken symmetry with incommensurate order parameters, and the Moir\'{e} superlattice in twisted van der Waals heteroustructures. Because of the incommensurability between the emergent superstructure and the underlying lattice, the crystal momentum is no longer a good quantum number, which invalidates the conventional band-structure description of electronic states. More importantly, incommensurability can have significant effects on the electron eigenstates. For instance, incommensurate potential can render the electronic wave functions localized or critical \cite{Siebesma_1987_Multifractal_Properties_of_Wave_Functions_for_One-Dimensional_Systems_with_an_Incommensurate_Potential}. This has been demonstrated in the Aubry-Andr\'{e} model, a canonical system for studying the incommensurability-induced electron localization, and its variants~\cite{Aubry_Analyticity_1980,PhysRevB.96.214201,PhysRevB.98.235116} and quasicrystal systems~\cite{PhysRevB.38.1621,PhysRevB.38.12903,PhysRevB.38.5981}. 

Collective electron behaviors are also expected to be modified by the presence of incommensurability, due to the altered nature of single-particle wavefunction. Indeed, superconductivity \cite{kamiya_discovery_2018} and unusual quantum critical state \cite{deguchi_quantum_2012} have been observed in quasicrystals. The recent experimental observation of superconductivity and correlated insulating states in the twisted bilayer graphene (TBG) with incommensurate structure is another example \cite{cao_correlated_2018,cao_unconventional_2018,Yankowitz1059,lu2019superconductors}. Although the single particle physics in TBG can be satisfactorily described by a continuum model neglecting the incommesurability of the Moir\'{e} pattern \cite{PhysRevB.86.155449,bistritzer_moire_2011}, the role of incommesurability on many body states remains unexplored \cite{PhysRevB.100.144202}. Motivated by these recent experimental progress, we study the superconductivity in a family of quasi-periodic systems both in one and two dimension.   

In quasi-periodic systems, the electronic states can be categorized into extend, localized, and critical states depending on the spatial characteristics of the wave functions. In the extended state, the wave function spreads extensively over the whole system even in the thermodynamic limit, and are analogous to the Bloch states in crystals. The localized state, on the other hand, exhibits a wave function that is confined to only a finite number of lattice sites. Most interestingly, a multifractal, self-similar structure emerges in the wave function of the critical state~\cite{HIRAMOTO_A_SCALING_APPROACH}. The different nature of these electron eigenstates also highlights a trade-off between the pairing strength and phase coherence of superconductivity. On one hand, although superconducting pairing can be maximized locally through confinement of electrons, superconductivity is disrupted due to the localized condensates. On the other hand, while a better phase coherence can be maintained by an extended wave function, delocalized electrons in such a state do not take full advantage of the short-range pairing interaction. As a consequence, the superconducting transition temperature $T_c$ is exponentially weak according to the BCS theory. This implies that $T_c$ may be enhanced in the case of critical states by optimizing the local pairing interaction while maintaining the long range phase coherence. This is indeed the case as will be revealed below.  

\section{Model}
We study a one dimensional $s$-wave superconductor with an incommensurate potential, described by a Hamiltonian~$H=H_0+H_{\text{sc}}$, with
\begin{gather}
\label{AA_Hamiltonian:NI}
H_0=-t\sum_{\langle ij \rangle,\sigma}c_{i\sigma}^{\dagger}c_{j\sigma}-\sum_{i,\sigma}\big(U_i+\mu\big)c_{i\sigma}^{\dagger}c_{i\sigma},\\
\label{AA_Hamiltonian:SC}
H_{\text{sc}}=-V\sum _i c_{i\uparrow}^{\dagger} c_{i\downarrow}^{\dagger} c_{i\downarrow} c_{i\uparrow}.
\end{gather}
Here $c_{i\sigma}^{\dagger}$ ($c_{i\sigma}$) is creation (annihilation) operator of electron with spin $\sigma$ on the $i$-th site of a periodic chain, $\mu$  is the chemical potential,  $U_i$ is the on-site potential, and $t$ is the nearest-neighbor hopping constant, which is set to $t=1$ for convenience in the following discussions. The Aubry-Andr\'{e} (AA) model corresponds to an incommensurate 
$U_i=J\cos(2\pi Qx_i)$, 
where $Q$ is an irrational number and $x_i$ is  position of the $i$-th site, so that the local potential becomes incommensurate with the underlying lattice. In this study, $Q$ is set to be the golden ratio, $(\sqrt{5}-1)/2$, and is approximated by the Fibonacci sequence $Q\approx F_{n-1}/F_n$, where $F_n$ is the $n$-th Fibonacci number. \SZL{We consider half filling by tuning $\mu$}. $H_{\text{sc}}$ describes the $s$-wave superconducting coupling and $V$ is the pairing strength. The AA model, described by $H_0$, exhibits a self-duality and a sharp localization-delocalization transition driven by $J$. It displays a spectrum consisting entirely of extended states for $J<2$, and of localized states for $J>2$. The quantum critical point $J = 2$ is characterized by a self-similar spectrum with all eigenstates becoming critical~\cite{Aubry_Andre_original}.

Standard Bogoliubov-de Gennes (BdG) method is used to solve this system \cite{Zhu_2016}.  The BdG Hamiltonian is
\begin{eqnarray}
H_{\mathrm{eff}}=H_{0}+\sum_{i}\left(\Delta_i c_{i\uparrow}^{\dagger} c_{i\downarrow}^{\dagger} + \Delta_{i}^{*} c_{i\uparrow} c_{i\downarrow}\right),
\end{eqnarray}
where $\Delta_i = V\langle c_{i\uparrow} c_{i\downarrow}\rangle$ is the local pairing amplitude. Here $H_{\mathrm{eff}}$ can be diagonalized by Bogoliubov transformation,
\begin{align}
c_{i \sigma }=\sum _{n}^{\prime} \left(u_{i \sigma }^n \gamma_{n} -\sigma v_{i \sigma }^{n*}\gamma_{n}^{\dagger }\right),\\
c_{i \sigma }^{\dagger }=\sum_{n}^{\prime}\left(u_{i \sigma }^{n*}\gamma_n^{\dagger }-\sigma v_{i \sigma }^n \gamma_{n} \right), 
\end{align}
where $\gamma^{\dagger}_n$ and $\gamma_n$ are the creation and annihilation operators for Bogoliubov quasiparticle at state $n$ and the prime sign means the sum is over all positive quasiparticle state $E_n>0$. The $u$ and $v$ coefficients are obtained from the BdG equations,
\begin{equation}
\sum_{j}
\begin{pmatrix}
h_{ij} & \Delta_i\\
\Delta_i^{*} & -h_{ij}^{*}
\end{pmatrix}
\begin{pmatrix}
u_{j\uparrow}\\v_{j\downarrow}
\end{pmatrix}
=E_{n}
\begin{pmatrix}
u_{i\uparrow}\\v_{i\downarrow}
\end{pmatrix},
\end{equation}
where
\begin{align}
    h_{ij}&=-t \delta_{\langle ij \rangle}-(J\cos(2\pi Qx_i)+\mu)\delta_{ij},\\
    \Delta_i&=\frac{V}{2}\sum_{n}u_{i\uparrow}^{n}v_{i\downarrow}^{n*}\tanh{\left(\frac{E_{n}}{2k_{B}T}\right)}.
\end{align}

\section{BdG results}
\begin{figure}[h]
    \includegraphics[width = \linewidth]{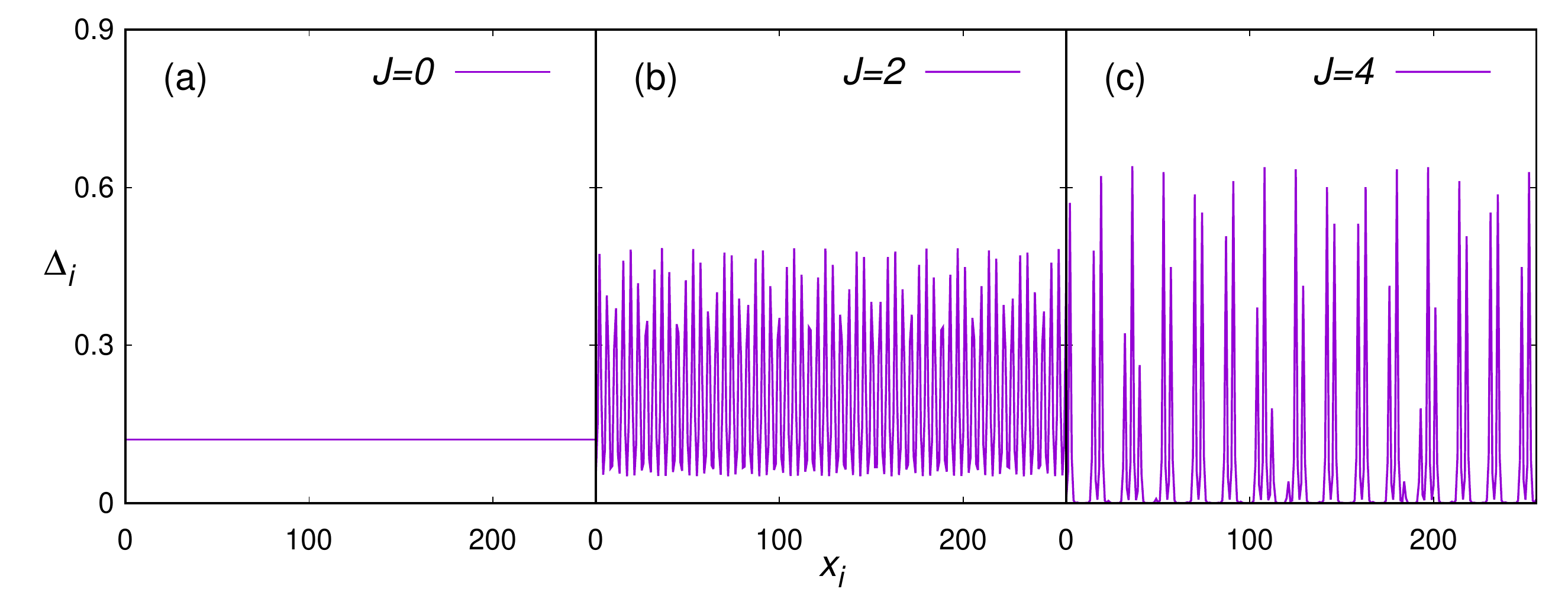}
    \includegraphics[width = \linewidth]{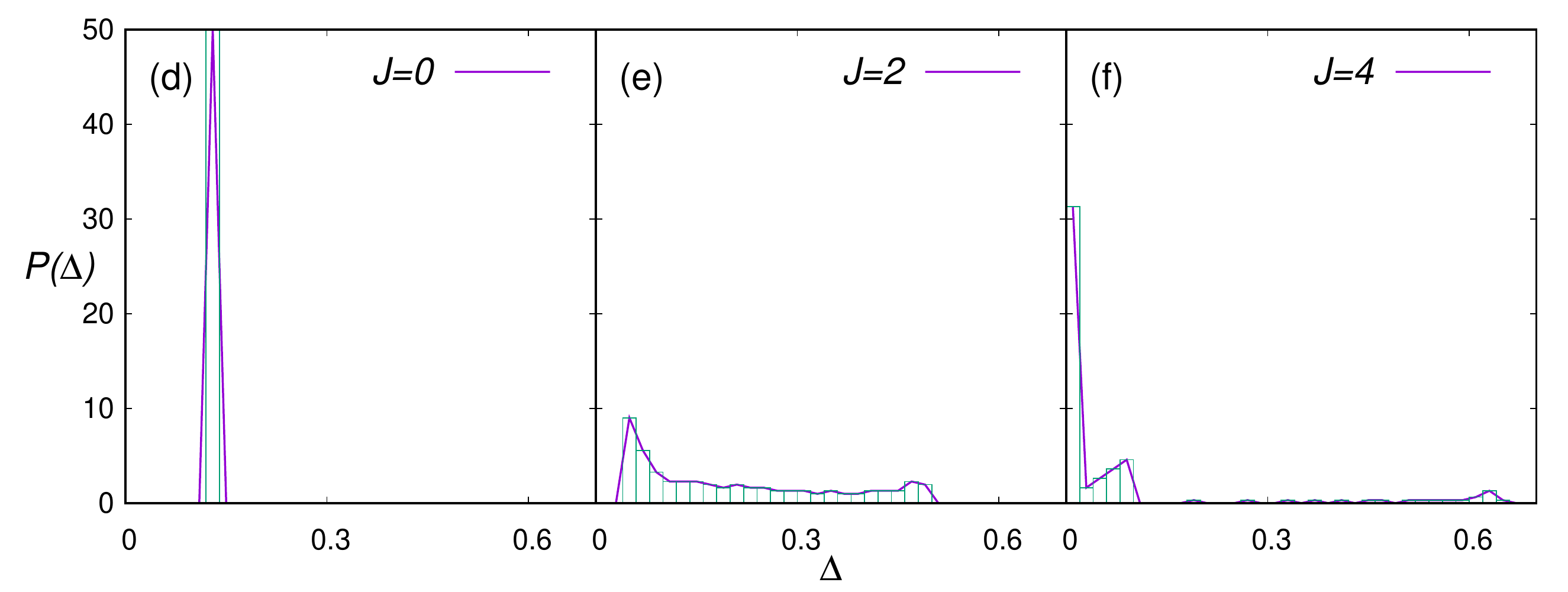}
    \includegraphics[width = \linewidth]{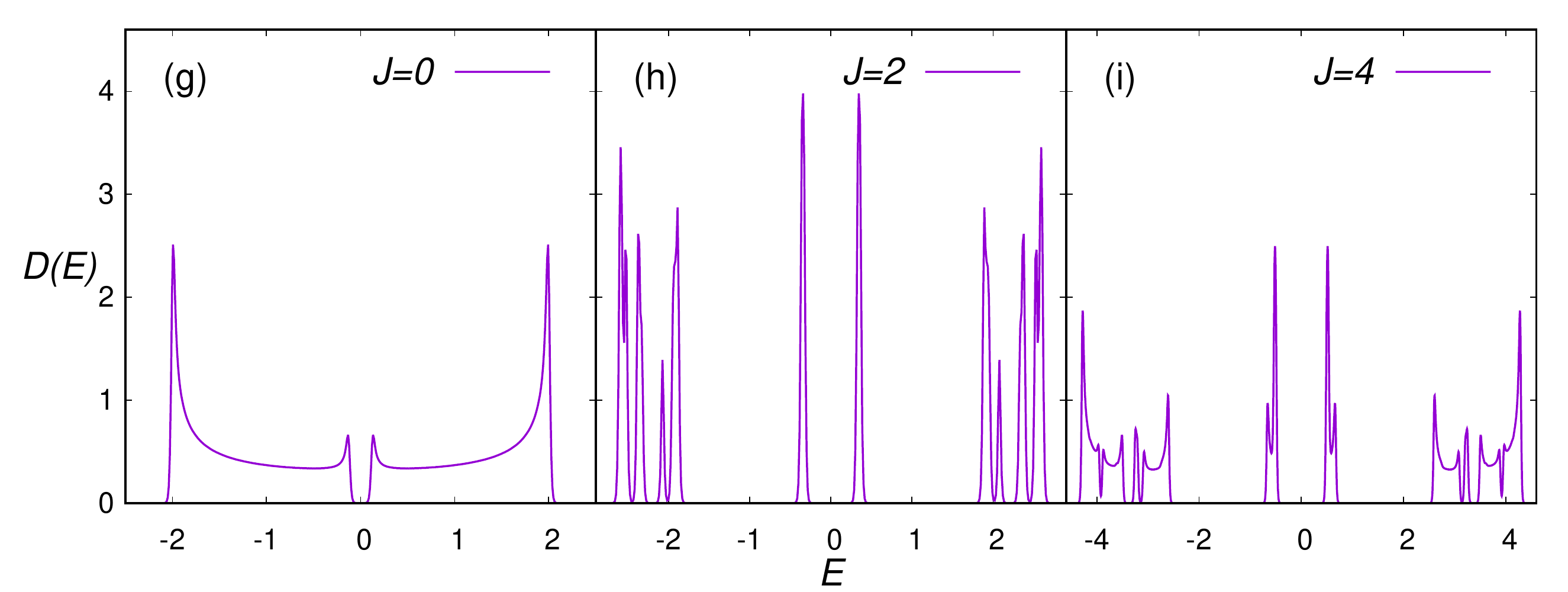}
    \caption{\label{Figure1_cp}
    BdG calculation results for a system of size $L=610$ with $V=1.5$ and $T=0.01$.
    (a)$\sim$(c) local order parameters as a function of position at $J=0,2,4$.
    (d)$\sim$(f) probability distribution function of local order parameter obtained using histogram method.
    (g)$\sim$(i) total density of states as a function of energy.
    }
    \label{Figure1}
\end{figure}

First we show the Bogoliubov-de Gennes (BdG) calculation results for the 1D $s$-wave superconductor under incommensurate potential $U_i=J\cos(2\pi Qx_i)$. The results of local order parameter, probability distribution of local order parameter and density of states, are shown in Fig.~\ref{Figure1}. It can be seen that for the extended states with $J=0$, the system is a standard homogeneous $s$-wave superconductor. When the system is critical at $J=2$, superconducting order parameter oscillates in space as evidenced by double peaks in the distribution $P(\Delta)$. In the localized region with $J=4$, there are superconducting islands with locally enhanced superconductivity separated by weak superconducting regions. In all cases, the spectrum is gapped around the chemical potential $E=0$.

Since the emergence of superconductivity requires the phase rigidity of the Cooper-pair condensates in an inhomogeneous state, here we use the superfluid stiffness $D_s$ to characterize the long-range phase coherence. It is given by~\cite{Zhu_2016,PhysRevB.47.7995,Nandini2001}
\begin{equation}
\frac{D_s}{\pi}=-\langle K_x \rangle + \Pi_{xx}(q\rightarrow 0, \omega=0),
\end{equation}
where $\langle K_x \rangle$ is the averaged kinetic energy and 
\begin{align}
    \Pi_{xx}(\mathbf{q}, \omega_n)=\frac{1}{N}\int_0^\beta d\tau \exp(i\omega_n\tau)\langle J_x(\mathbf{q}, \tau)J_x(-\mathbf{q}, 0)\rangle,
\end{align}
is the retarded correlation function of the particle current operator, 
\begin{align}
    J_x(\mathbf{q})=it \sum_l \exp(-i \mathbf{q}\cdot\mathbf{x}_l)(c_{l+s,\sigma}^\dagger c_{l,\sigma}-c_{l,\sigma}^\dagger c_{l+s,\sigma}).
\end{align}
The details on the calculation of $D_s$ based on the BdG method is presented in Appendix A. In the well localized phase, the global phase coherence is established by a weak Josephson type coupling between strong superconducting islands separated by weak superconducting regions~\cite{PhysRevB.32.5658}. The energy of the superconducting condensate can be approximated as $E\propto -D_s\sum_{\langle i,j\rangle}\cos(\theta_i-\theta_j)$, where $\theta_i$ is the phase of the superconducting order parameter at $i$-th strong superconducting island \cite{Nandini2001}. $D_s$ increases with the amplitude of the superconducting order parameter in the strong superconducting islands and the overlap of the order parameter between these islands.

\begin{figure}
    \centering
    \includegraphics[width = 1.0\linewidth]{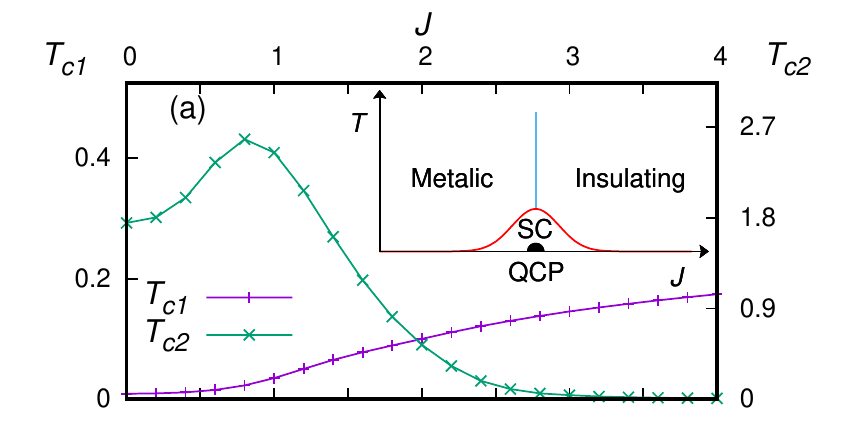}
    \includegraphics[width = 1.0\linewidth]{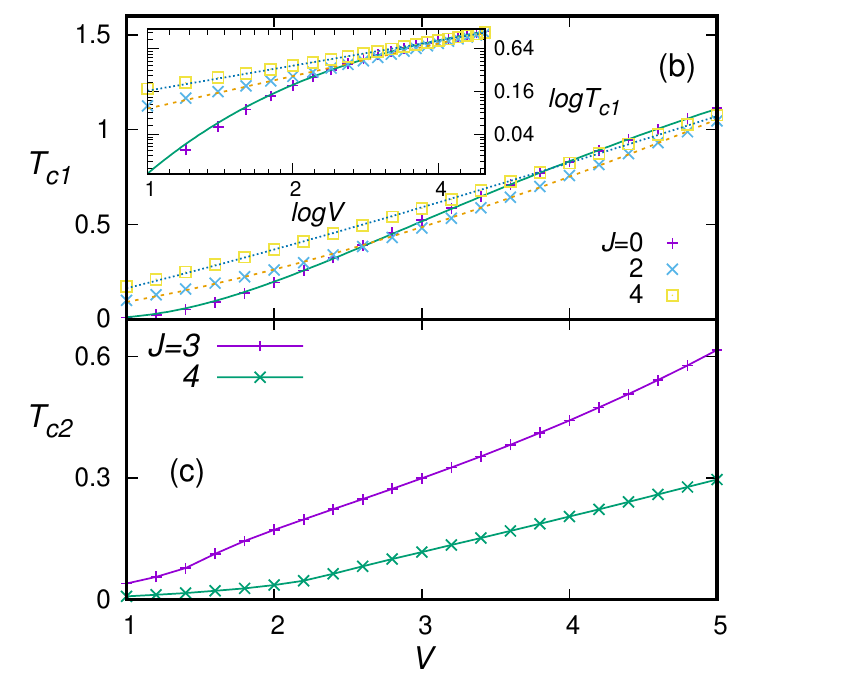}
    \caption{(a) Critical temperature and zero temperature superfluid stiffness as a function of $J$ at $V=1$. $T_{c1}$ determined from the averaged amplitude of the order parameter is enhanced as $J$ increase while $T_{c2}\equiv D_s(T=0)$ decrease exponentially as the system enters localized region. 
    Inset is a schematic phase diagram based on (a), which shows the existence of a superconducting (SC) dome near the localization quantum critical point (QCP).
    (b) $T_{c1}$ vs $V$ for $J=0,\ 2,\ 4$. The numerical results can be fitted by $T_{c1}\propto e^{-0.17/V}$  for $J=0$, $T_{c1}\propto V^{1.6}$ for $J=2$ and $T_{c1}\propto V^{1.1}$ for $J=4$. The inset plot is the log-log plot to illustrate the power law behavior.
    (c) $T_{c2}$ vs $V$ in the localized region, $J=3$ and $J=4$.
    }    
    \label{fig:AA_BdG_results}
\end{figure}

Although strictly speaking, there is no long-range superconductivity order in 1D, our mean-field approach to the superconducting AA model should be viewed as a quasi-1D approximation to either 2D or 3D incommensurate superconductivity. (A true 2D model calculation will be presented in Sec. \ref{sec6}) With this understanding, superconductivity is destroyed by suppressing either the amplitude of the order parameter or the phase coherence. In the extended state as in the case of conventional BCS theory, $T_c$ is limited by the averaged amplitude of the order parameter over the whole system. In the localized state, $T_c$ is limited by phase fluctuation and is proportional to the zero temperature $D_s$, which measures the coupling between different superconducting islands. Here we define two temperature scales: $T_{c1}$ is the temperature when $\Delta$ vanishes throughout the system, and $T_{c2}=D_s(T=0)$ represents the energy scale of phase coherence. And we estimate the transition temperature $T_c$ of our system by min($T_{c1}$, $T_{c2}$). \SZL{As $J$ increases, the system becomes more spatially localized. $T_{c1}$ increases because the system can take the advantage of local pairing interaction. Meanwhile $T_{c2}$ diminishes as the superconductivity becomes more spatially localized. Therefore the superconductivity is limited by $T_{c2}$ for a strong incommensurate potential. }

BdG calculation results of $T_{c1}$ and $T_{c2}$ of a system with size $L=233$ are presented in Fig.~\ref{fig:AA_BdG_results}. {In Appendix D, We check that the finite size effect is negligible by comparing to the results with a larger $L$.} $T_{c1}$ increases monotonically with $J$. In the localized state, $T_{c1}$ corresponds to the highest transition temperature among all superconducting islands. As the electronic wave functions become more localized by increasing $J$, the electrons can take full advantage of the local pairing interaction, and as a consequence, local superconductivity is enhanced. On the other hand, $T_{c2}$ is first enhanced with increasing~$J$ until a critical $J_c$, whose origin is unclear. As $J$ is further increased, $T_{c2}$ starts to decrease due to the loss of phase coherence between spatially localized superconducting islands. The dependence of $T_{c1}$ and $T_{c2}$ on $J$ indicates the existence of a superconducting dome near the localization transition at $J=2$, as schematically depicted in the inset of Fig.~\ref{fig:AA_BdG_results}(a). It is also worth pointing out the different nature of superconducting transition on the two sides of the dome. In the extended regime corresponding to small $J$, the system undergoes a superconductivity to metal transition due to a vanishing amplitude of Cooper pairing upon increasing temperature. In the localized phase, there is a temperature-driven superconductor-to-insulator transition caused by the phase fluctuations of the superconducting order parameter. At finite temperature, there is no sharp distinction between metals and insulators and we expect a smooth crossover tween metallic and insulating state for temperature above $T_c$.

The dependence of $T_{c1}$ and $T_{c2}$ on $V$ are plotted in Fig.~\ref{fig:AA_BdG_results}(b) for the three different types of electron eigenstates in AA model. For extended wave function, $T_{c1}$ has a standard BCS exponential relation with $1/V$. In the localized state, the relation between $T_{c1}$ and $V$ is almost linear. Interestingly, at the critical point $J = 2$, $T_{c1}$ increases with $V$ according to a power law: $T_{c1}\propto V^{1.6}$. On the other hand, $T_{c2}\propto \exp(-1/a V)$ in the localized state. Therefore in the weak coupling limit $V\ll t$, $T_{c}$ is exponentially weak in $1/V$ both in the localized and extended regions, while $T_c$ is enhanced significantly near the localization transition as it depends on $V$ by a power law.

In the AA model, the electronic spectrum form bands for the extended states. Both for the critical and localized states, instead of form bands, the spectrum is point-like~\cite{PhysRevLett.51.1198_Fibonacci_Model}. Therefore, it is likely that the chemical potential locates in the gap of the single-particle spectrum, and therefore a threshold $V$ is required to trigger superconductivity.

\section{Weak coupling theory}

The BdG method is restricted to a large $V$ because the superconducting coherence length $\xi$ increases exponentially fast for a weak $V$ [$\xi \sim \exp(1/N_0 V)$ with $N_0$ the density of state]. This would require large system size $L\gg \xi$, which is practically impossible. \SZL{To reach the weak coupling limit $V/t\ll1$ and also to understand the dependence of $T_{c1}$ on $V$, we provide analytical description based on Anderson's idea of pairing the time-revered  eigenstates of the non-interacting system~\cite{ANDERSON195926,Nandini2001}.} The non-interacting time-reversal symmetric Hamiltonian $H_0$ is bilinear and can be exactly diagonalized: $H_0|\psi_{\alpha\sigma}\rangle=\epsilon_\alpha|\psi_{\alpha\sigma}\rangle$, where $\alpha$ labels the exact eigenstates of $H_0$. We rewrite $H$ in this basis, $d^\dagger_{\alpha\sigma}\ket{0}=\ket{\psi_{\alpha\sigma}}$ and only consider the pairing interaction between time-reversed states, $\ket{\psi_{\alpha\uparrow}}$ and $\ket{\psi_{\bar{\alpha}\downarrow}}$:
\begin{equation}
\label{M_matrix:Hamiltonian}
H^{\prime}=\sum_{\alpha,\sigma}\epsilon_{\alpha}d_{\alpha\sigma}^{\dagger}d_{\alpha\sigma}
-V\sum_{\alpha,\beta}M_{\alpha\beta}d_{\alpha\uparrow}^{\dagger}d_{\bar{\alpha}\downarrow}^{\dagger}d_{\bar{\beta}\downarrow}d_{\beta\uparrow},
\end{equation}
where
\begin{align}
    M_{\alpha\beta}=\int\psi_\alpha^*\psi_{\bar{\alpha}}^*\psi_\beta\psi_{\bar{\beta}}\mathrm{d}r.
\end{align}
Here $\psi_{\bar{\alpha}}$ is the time reversal partner of $\psi_\alpha$. The linearized gap equation for $\Delta_\beta=V\langle c_{\bar{\beta}\downarrow}c_{\beta\uparrow}\rangle$ at temperature $T_{c1}$ is
\begin{equation}
\label{Gap_equation}
\Delta_\alpha=V\sum_\beta M_{\alpha\beta}\frac{\Delta_\beta}{2\epsilon_\beta}\tanh\left(\frac{\epsilon_\beta}{2T_{c1}}\right),
\end{equation} The characteristic of the normal state electronic wave function is contained in the $M$ matrix.

For the extended states $J<2$, wave functions extend over the entire lattice and the amplitudes of wave functions scale as the inverse square root of the system size $L$, $|\psi_\alpha|\propto 1/\sqrt{L}$. Thus, $M_{\alpha\beta}$ is independent of $\alpha$, $\beta$ and scales as $1/L$, which leads to a gap equation with the standard BCS form and therefore $T_{c1}  \propto e^{-1/aV}$. (see Appendix B for detailed calculations) For the localized states, the wave functions are confined in small regions characterized by a localization length $\xi_l$ and they scale as $|\psi_\alpha|\propto 1/\sqrt{\xi_l}$. The wave function has negligible overlap with wave functions of other states. As a result, only the diagonal terms of $M_{\alpha \beta}$ are important, $M_{\alpha \beta}=\delta_{\alpha \beta}/\xi_l$, which results in a linear dependence of $T_{c1}$ on $V$, $T_{c1} \propto a V$. The results of $T_{c1}$ vs $V$ at $J=4$ is shown in Fig.~\ref{fig:AA_BdG_results} (b). The dependence of $T_{c1}$ on $V$ deviates slightly from a linear behavior because of the nonzero overlap of wave functions at different energies when $J$ is not large.

\begin{figure}
    \centering
    \includegraphics[width = 0.95\linewidth]{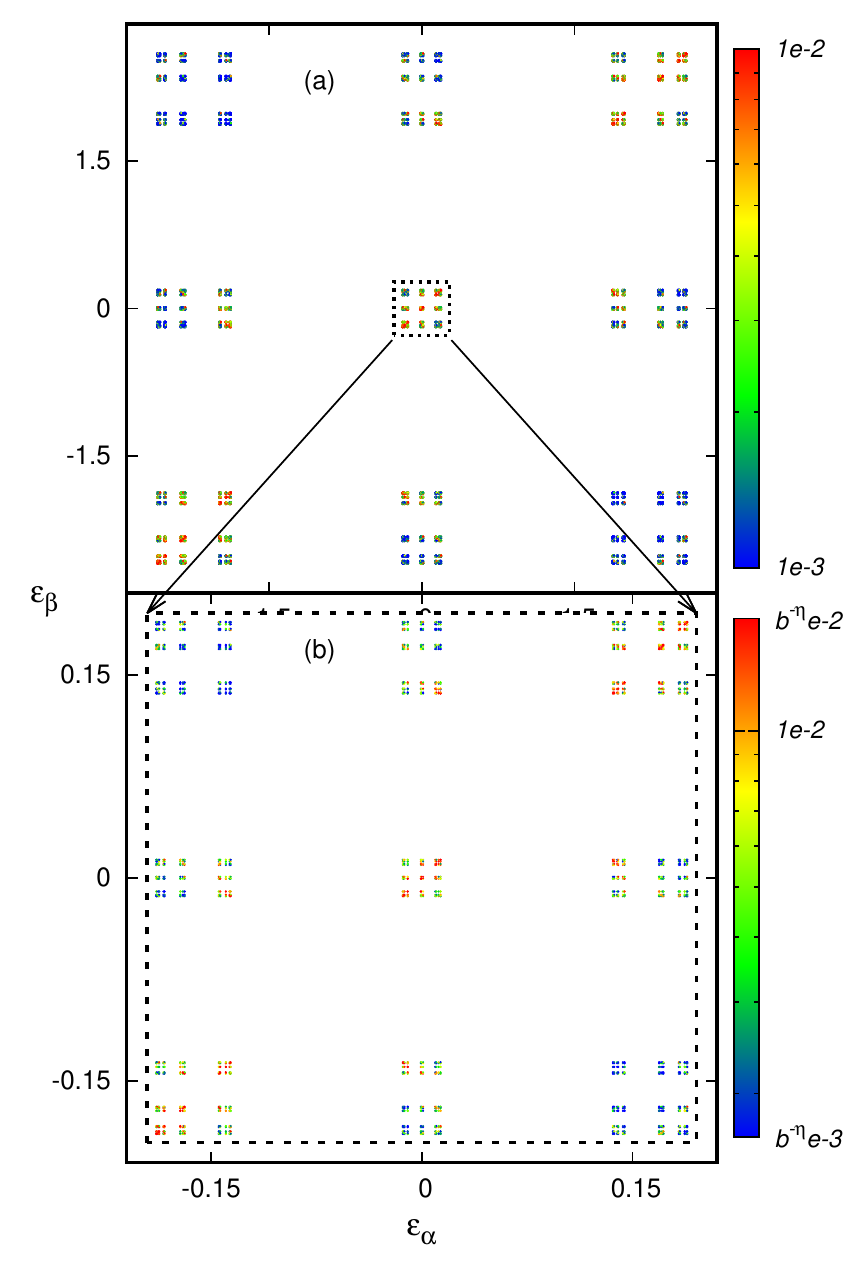}
    \caption{(a) $M_{\alpha\beta}$ plotted in full range of energy. The matrix element $M_{\alpha\beta}$ is plotted as a dot at $x=\varepsilon_\alpha$, $y=\varepsilon_\beta$ and the magnitude of matrix element is indicated by the color of the dot. (b) Zoomed in with a factor of $b=13.8$. The structure of the zoomed matrix is almost identical to the original indicating that $M_{\alpha\beta}$ is self-similar at different energy scale. 
    }
    \label{fig:M_matrix}
\end{figure}

In critical state $J=2$, the spectrum is self-similar, which is characterized by a multifractal  exponent $\alpha_M$ and its distribution $f_M(\alpha_M)$~\cite{PhysRevB.34.2041}. Therefore one would also expect $M_{\alpha \beta}$ to be self-similar, i.e. $M_{\alpha \beta}\equiv M(\epsilon_\alpha, \epsilon_\beta)=b^{-\eta} M(\epsilon_\alpha/b, \epsilon_\beta/b)$. For simplicity, we have assumed that $M$ is characterized by a single exponent $\eta$. Because the spectrum is discrete, this scaling transformation is valid only for discrete value of $b$, as shown in Fig.~\ref{fig:M_matrix}. The scaling property of $M$ immediately leads to a power-law relation between $T_c$ and $V$. If the superconducting coupling strength $V$ is scaled by a factor $\alpha$, $V\rightarrow \alpha V$, the energy level $\epsilon$ must also be scaled by a factor $b$, $\epsilon\rightarrow\ b \epsilon$, to maintain the form of the gap equation unchanged. The term in $\tanh$ is dimensionless, thus the temperature $T$ must also be scaled by $b$. From this scaling argument, we can obtain the power-law dependence $T_{c1} \propto V^{1/(1+\eta)}$. 

The spectrum of AA model is self-similar at $J=2$. At $J=2$, the spectrum does not have any continuous bands and for a finite system there is no one-to-one correspondence between the original $M$ matrix and the zoomed one. Therefore, the value of $\eta$ is estimated by calculating the self-similar scaling of largest elements in each small blocks of $M_{\alpha\beta}$. The scaling exponent of each block is then averaged to obtain the final $\eta$.  For  $L=1597$ and $L=6765$, the matrix can  be rescaled by a scaling factor $b\approx13.8$ and the estimated exponent is $\eta\approx-0.19$ which gives $T_{c1}\propto V^{1.23}$. The scaling analysis agrees reasonably well with the numerical fitting result $T_{c1}\propto V^{1.6}$. The slight deviation in the exponent could be caused by the finite size effect because extremely large system size is required to capture the self similarity behavior of $M_{\alpha \beta}$ with a rescaling factor $b\approx13.8$. The single exponent approximation to the scaling relation for the $M_{\alpha \beta}$ can also cause deviation. \SZL{Now it becomes clear from Eqs. \eqref{M_matrix:Hamiltonian} and \eqref{Gap_equation} that the fractal nature of normal state wave function renders the effective pairing interaction being fractal. As a consequence, $T_c$ depends on the bare pairing interaction by a power law function, and superconductivity is enhanced.}

\section{Other 1D models}
\begin{figure}[t]
    \centering
    \hspace*{-0.8cm}\includegraphics[width = 0.95\linewidth]{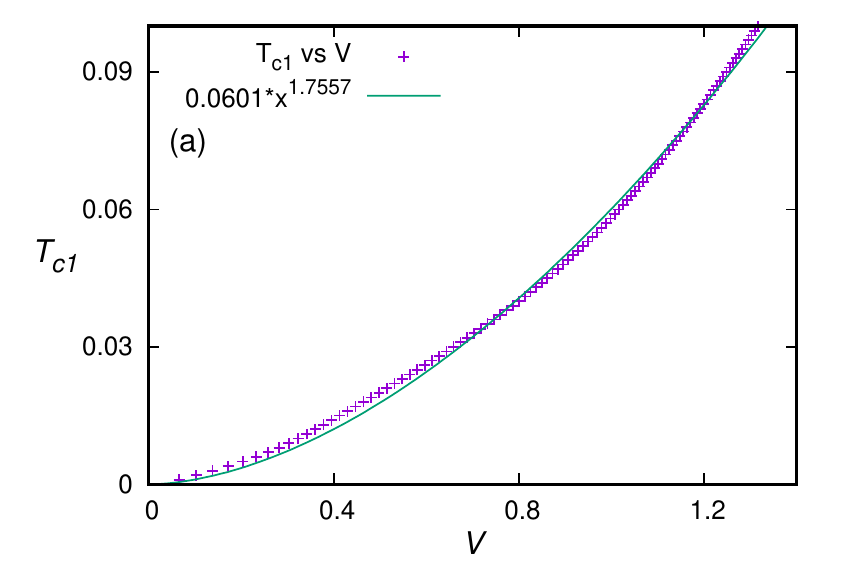}
    \hspace*{-0.8cm}\includegraphics[width = 0.95\linewidth]{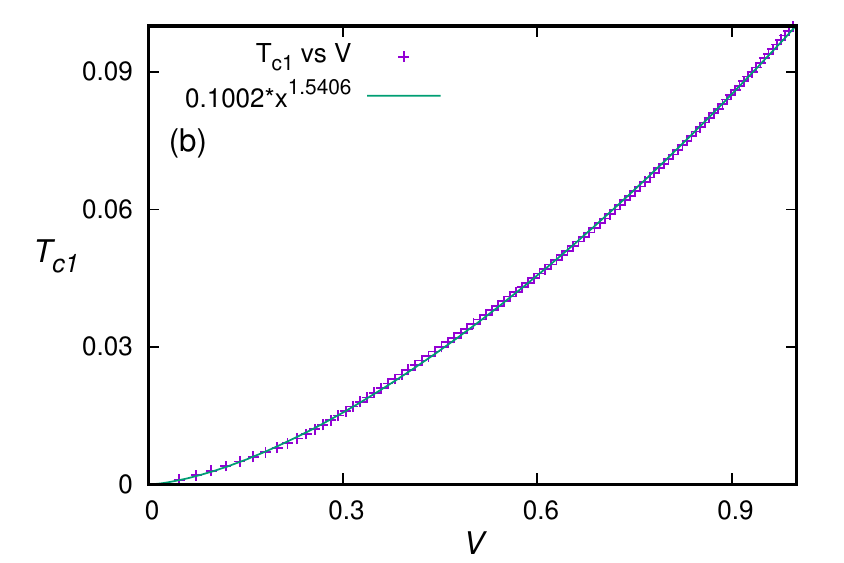}
    \caption{$T_{c1}$ vs $V$ calculated using the $M$ matrix method with $J=1$, (a) $\mu=-0.4293$, (b) $\mu=2.06156$ for the Fibonacci model with $L=987$. Symbols are numerical results and lines are fitting curves.}
    \label{fig:Fibonacci_Tc_vs_V}
\end{figure}

The power law dependence of $T_{c1}$ on $V$ is due to the self-similarity of $M_{\alpha\beta}$, which can be demonstrated in the Fibonacci model with an onsite incommensurate potential given by $U_i=U(Qx_i)$ and $U(x)=-J$ for  $m-Q\leqslant x \leqslant m$, $J$ for $m<x<m+1-Q$, where $m$ is an arbitrary integer.
The electronic spectrum is always critical regardless the strength of the potential \cite{PhysRevLett.50.1873_Fibonacci_Model,PhysRevLett.51.1198_Fibonacci_Model,PhysRevLett.50.1870_Fibonacci_Model,HIRAMOTO_A_SCALING_APPROACH}, see also Appendix C. It can be seen from Fig.~\ref{fig:Fibonacci_Tc_vs_V} that in this model, $T_c$ and $V$ always has a power law relation provided that the Fermi level is not in a gap of the non-interacting spectrum.

\begin{figure}
    \centering
    \includegraphics[width = 1.0\linewidth]{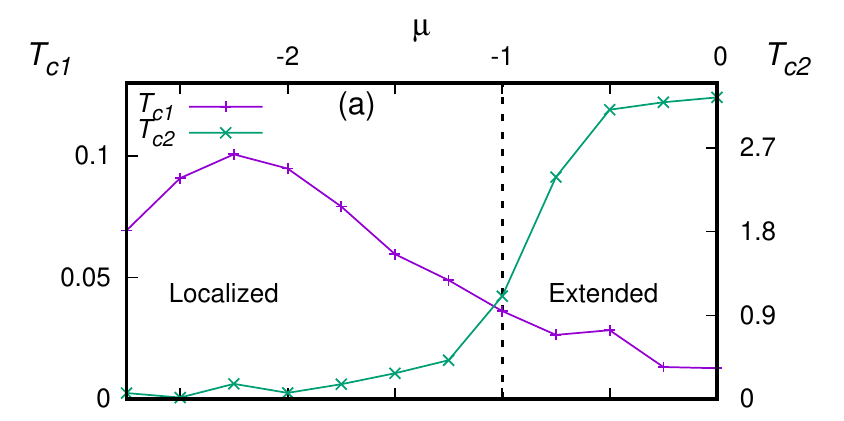}
    \includegraphics[width = 1.0\linewidth]{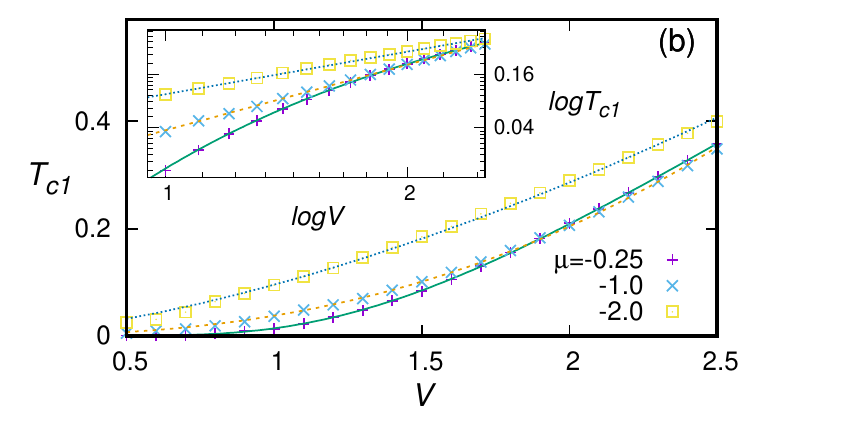}
    \caption{(a) 
    Critical temperature and zero temperature superfluid stiffness as a function of $\mu$ with $V=1$, $J=1$, $\nu=0.7$. The mobility edges are at $\mu=\pm1.0$, marked by the dotted line in the plot. A superconducting dome exists near the mobility edge $\mu=-1.0$. (b) $T_c$ vs $V$ for $\mu=-0.25,\ -1.0,\ -2.0$. For $\mu=-0.25$, $\mu=-1.0$ and $\mu=-2.0$, the numerical data can be described by $T_c\propto e^{-0.18/V}$, $T_c\propto V^{2.4}$, $T_c\propto V^{1.6}$ respectively. The inset plot is the log-log plot.
    }
    \label{fig:GA_BdG_results}
\end{figure}

In the AA model, all eigenstates have the same spatial characteristics, and the localization-delocalization transition is controlled by the strength $J$ of the incommensurability. It is also possible in certain class of incommensurate models that the localized and extended states coexists in the spectrum and are separated by a mobility edge, at which the wave functions become critical. One can thus change the wave function characteristics by tuning the chemical potential $\mu$. Therefore there can exist a superconducting dome as a function of $\mu$ near the mobility edge. We demonstrate this scenario explicitly using the generalized Harper model with a modulated incommensurate potential $U_i=J\cos(2\pi Q x_i^\nu)$. Without superconducting coupling, this model exhibit a continuous spectrum with mobility edges at $\pm(2-J)$ for $0<\nu<1$ and $J<2$ \cite{PhysRevLett.61.2144_modulated_harper_model}, see also Appendix C. 
The states at the band edge are localized and the states in the middle of the band are extended. The BdG results of a system of $L=987$  show a enhancement of superconductivity near the mobility edge forming a superconducting dome, Fig~\ref{fig:GA_BdG_results}. When $\mu=-1.0$ at $J=1$, superconductivity is mainly contributed from the states near the mobility edge which are critical. Hence, there is a power law relation between $T_c$ and $V$ at $\mu=-1.0$. Superconductivity is suppressed when the chemical potential is tuned to localized region. Our BdG results for $\mu=-2.0$ show that $T_{c1}$ depends on $V$ by a power law with a smaller exponent. This power law dependence is due to the relatively large $V$ required by the BdG calculations, where both the localized state and critical states contribute to superconductivity. In the weak coupling limit when only localized states participate in the pairing, $T_{c1}$ scales with $V$ linearly according to Eq.~\eqref{Gap_equation}.

\begin{figure}[h]
    \centering
    \includegraphics[width = \linewidth]{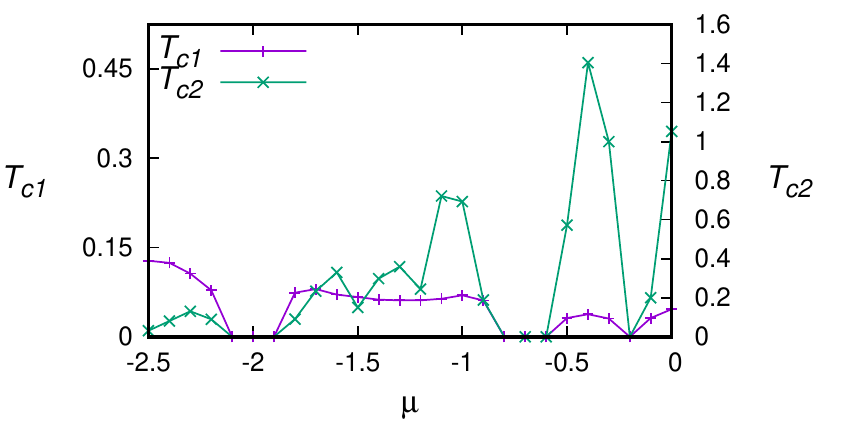}
    \caption{$T_{c1}$ and $T_{c2}$ vs $\mu$ for the double cosine potential $L=377$ with $J_1=0.75$ and $J_3=0.75$. There is no superconducting dome because of the absence of the critical states when $\mu$ is swept.}
    \label{fig:DoubleCosine}
\end{figure}

We next consider a model where the localized and extended states are separated by an energy gap in the spectrum. This is realized using the double cosine potential $U_i=J_1\cos(2\pi Qx_i)+J_3\cos(6\pi Qx_i)$ \cite{HIRAMOTO_A_SCALING_APPROACH}. In this case, no critical state exists, and the dependence of $T_{c1}$ and $V$ follows either that for the extended states or for localized states as shown in Fig.~\ref{fig:DoubleCosine}.

\section{2D AA model}\label{sec6}
To further support our conclusion in 1D, we perform additional calculations in 2D. The 2D system also allows one to study the superconducting transition in the presence of thermal fluctuations. We consider $s$-wave superconductivity in a generalized Aubry-André model in 2D with an incommensurate potential
\begin{align}
U_i=-J\sum _{i,\sigma } \left[\cos (2 \pi (x_i+y_i) Q) + \cos  (2 \pi (x_i-y_i) Q) \right] c_{i \sigma }^{\dagger} c_{i \sigma }.  \nonumber
\end{align}
Here, $x_i$ and $y_i$ are 2D coordinates of the $i$-th site.

\begin{figure}[b]
    \centering
    \hspace*{-0.8cm}\includegraphics[width =0.95\linewidth]{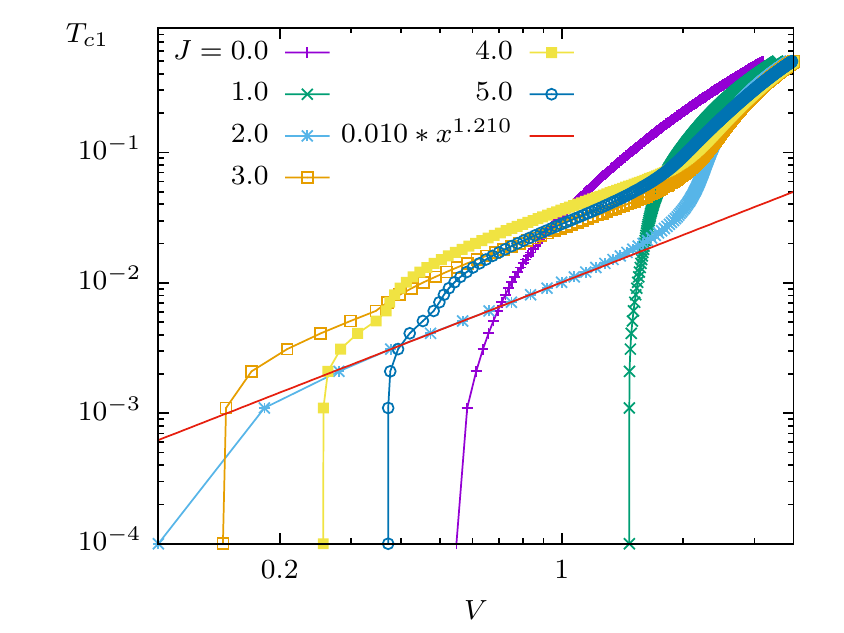}
    \caption{$T_c$ vs $V$ calculated based on weak coupling theory for a generalized 2D Aubry-André model. The system size is $L\times L=89\times 89$. The system is at half-filling with the chemical potential $\mu=0$ due to the symmetric spectrum of the system.  }
    \label{fig:2D_m_matrix_Tc_vs_V}
\end{figure}

The localization transition occurs at $J=2 $ when superconductivity is absent \cite{PhysRevB.101.014205}. We calculate $T_c$ using the weak coupling theory described in Eqs. \eqref{M_matrix:Hamiltonian} and \eqref{Gap_equation}. As shown in Fig.~\ref{fig:2D_m_matrix_Tc_vs_V}, at the localization transition, $T_c \propto V^{1.2}$. The power law dependence of $T_c$ on $V$ means an enhancement of $T_c$ in comparison to the standard weak coupling BCS theory results for a uniform system. We note that $T_{c1}$ drops to zero quickly for a very small $V$. This is because of the finite size effect, when $V$ is comparable to the discrete single particle spectrum gap. The results for a large $V$ deviate from the power law behavior because the weak coupling approximation in Eqs. \eqref{M_matrix:Hamiltonian} and \eqref{Gap_equation} breaks down.

\section{Effect of Coulomb interaction}
In the critical or localized states, the effect of Coulomb interaction is also enhanced \cite{bulaevskii_anderson_1985,SADOVSKII1997225}, similar to the pairing interaction. The effect of Coulomb interaction can be introduced by an energy dependent pairing strength $V_{\alpha \beta }=V \left(\epsilon _{\alpha }-\epsilon _{\beta }\right)$ \cite{KettersonBook1999}
\begin{equation}
   V \left(\epsilon _{\alpha }-\epsilon _{\beta }\right)=
   \left\{ 
   \begin{array}{cc} 
   V_p-V_c & \left| \epsilon _{\alpha }-\epsilon _{\beta }\right| \leq \hbar \omega _D\\
   -V_c & \hbar \omega _D<\left| \epsilon _{\alpha }-\epsilon _{\beta }\right| \leq \hbar \omega _c\\
   0 & \hbar \omega _c<\left| \epsilon _{\alpha }-\epsilon _{\beta }\right|  
   \end{array} 
   \right.
\end{equation}
where $\omega_D$ is the Debye frequency and $\omega_c$ is the frequency associated with the Coulomb interaction. $V_p<0$ is the attractive interaction and $V_c>0$ is the repulsive Coulomb interaction.

\begin{figure}[ht]
    \centering
    \hspace*{-0.8cm}\includegraphics[width=0.95\linewidth]{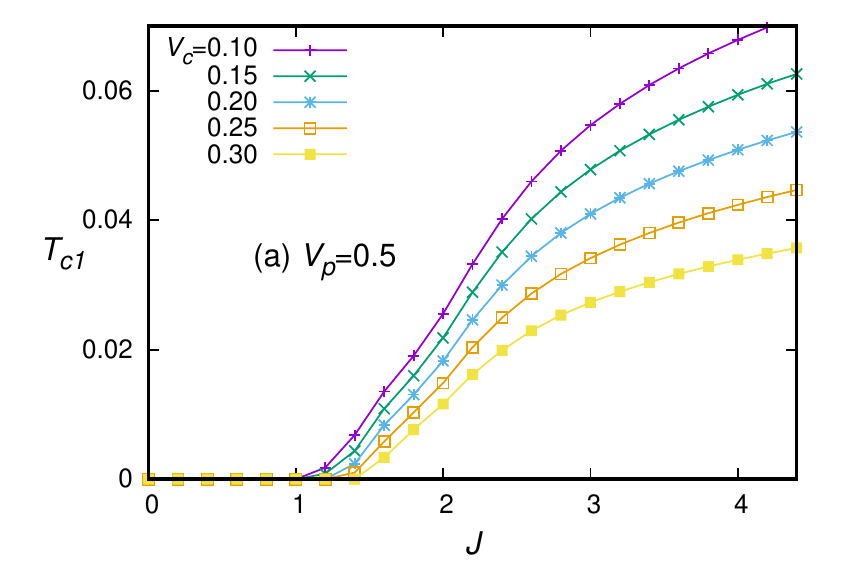}
    \hspace*{-0.8cm}\includegraphics[width=0.95\linewidth]{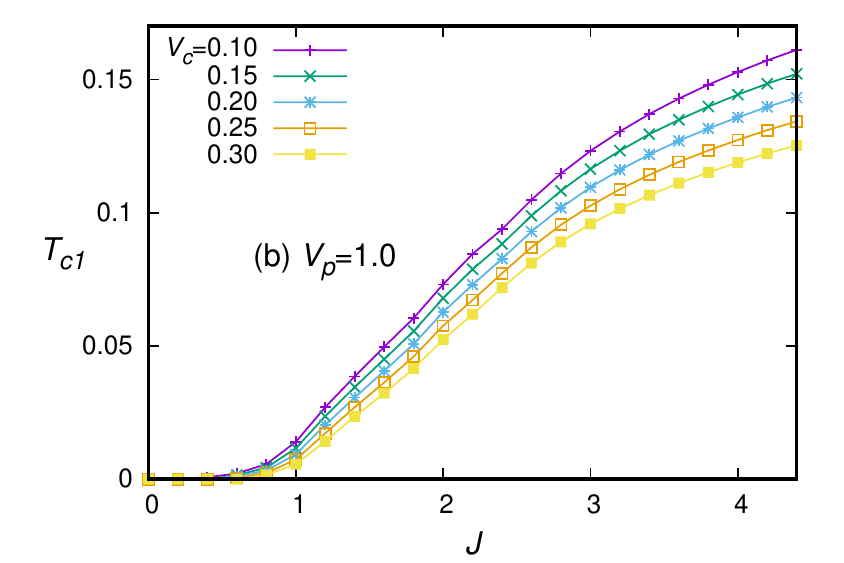}
    \caption{$T_{c1}$ vs $J$ of the superconducting AA model with Coulomb interaction for system size $L=987$ with several $V_{c}$s at (a) $V_p=0.5$ and (b) $V_p=1.0$.}
    \label{fig:Coulomb}
\end{figure}

We take \(\hbar \omega _D=0.3\), \(\hbar \omega _c=0.5\). The results of $T_{c1}$ vs $J$ of superconducting AA model with Coulomb interaction are shown in Fig.~\ref{fig:Coulomb}. $T_{c1}$ for a given $J$ is suppressed by the Coulomb interaction, and it is enhanced when the system is tuned to the more localized side at a given $V_c$ by increasing $J$. Note that $T_{c1}$ is determined from the amplitude of the order parameter. When the system enters the localized region $J>2$, the superconductivity is limited by superfluid stiffness, which is suppressed due to localization. Thus, a superconducting dome around the localization transition is expected even in the presence of Coulomb interaction.

\section{Discussion and Conclusions}
{As shown in Appendix E, the normal state density of state (DOS) at Fermi energy is modified by the incommensurate potential, and hence affects $T_c$. However, the change of DOS cannot explain the power law dependence of $T_c$ on the pairing interaction. So far, we have mainly focused on the effect of an incommensurate potential on $T_c$, which is determined by normal state wave functions and spectrum. When superconductivity is fully developed far below $T_c$, superconductivity can affect the localization transition by gapping the quasiparticle energy spectrum. It is possible that the localization transition at $T=0$ is completely masked by superconductivity.     }

Let us discuss the relation of our work to others. Similar phenomenology has been discussed in superconductors with random disorders, where a power law dependence of $T_c$ on $V$ is found based on scaling analysis near the localization transition~\cite{Fractal_superconductivity,PhysRevLett.98.027001,PhysRevLett.108.017002}. The enhancement of $T_c$ by disorders due to the multifractal electronic state was studied theoretically \cite{PhysRevB.92.174526} and observed in experiments \cite{zhao_disorder-induced_2019}. The effect of random disorder on superconductivity was studied by solving the BdG equation numerically in 2D. No enhancement of superconductivity was found because of the absence of localization transition in the standard Anderson model in 2D. \cite{PhysRevLett.81.3940,Nandini2001} A different mechanism for the enhancement of $T_c$ due to the enhancement of density of state by disorders was studied in Ref. \cite{PhysRevB.98.184510}. In Ref. \cite{PhysRevB.72.060502}, it is argued that impurities can cause spatial modulation in the pairing potential, which enhances $T_c$. In these cases, the single particle spectrum is continuous, which is different from that in the AA model studied here. We remark that the quasi-periodic potential has weaker effect on the localization of electronic wave function than the random disorders. This allows us to study the enhancement of the superconductivity near the localization transition or mobility edge in 1D models, which is not possible for random disorders, see Appendix F for more detailed discussions.

To summarize, we study the effect of incommensurate potential on 1D $s$-wave superconductors and found an enhancement of superconductivity near the localization transition in a class of quasi-periodic crystals. At the localization transition, $T_c$ depends $V$ by a power law, which gives rise to a superconducting dome near the localization critical point. In the region with extended states, superconductivity is destroyed by the suppression of the amplitude of the superconducting order parameter; while in the localized states, superconductivity is killed by the fluctuations of the phase of the superconducting order parameter. Our results suggest a promising routine to enhance $T_c$ of superconductors by incommensurate potentials.

\begin{acknowledgments}
The authors thank Alexander V. Balatsky, Ivar Martin, Qimiao Si and Senthil Todadri for helpful discussions.  Computer resources for numerical calculations were supported by the Institutional Computing Program at LANL.
This work was carried out under the auspices of the U.S. DOE Award No. DE-AC52-06NA25396 through the LDRD program, and was supported by the Center for Nonlinear Studies at LANL.
{This work is partially supported by the Center for Materials Theory as a part of the Computational Materials Science (CMS) program, funded by the U.S. Department of Energy, Office of Science, Basic Energy Sciences, Materials Sciences and Engineering Division. The authors also acknowledge the support of Advanced Research Computing Services at the University of Virginia.}
\end{acknowledgments}

\begin{widetext}
\begin{appendix}

\section{Derivation and calculation of superfluid stiffness}
As explained in the main text, the superfluid stiffness of the system is necessary for the estimation of critical temperature $T_{c2}$ in the localized region. Here shows the derivation of superfluid stiffness expressed in the BdG framework.
The derivation based on the method used in Ref. \cite{Zhu_2016}.
Consider a general Hamiltonian,
\begin{equation}
H=\sum _{i,j,\sigma } c_{i \sigma }^{\dagger }\left[-t_{i j}-\left(\mu +\frac{U}{2}-U_i^{\text{imp}}\right)\delta _{\text{\textit{$i j$}}}\right]c_{j
\sigma }+U\sum _i n_{i\uparrow }n_{i\downarrow }-\frac{V}{2}\sum _{i\neq j} n_in_j.
\end{equation}
We consider short range hopping.
The particle current and the local kinetic energy associated with the $x$-oriented hopping can be written as,
\begin{equation}
J_x^P\left(\pmb{r}_i\right)=i\sum _{\sigma } \sum _{j>i} \frac{\left(x_j-x_i\right)}{a}t_{i j}\left(c_{j \sigma }^{\dagger }c_{i \sigma }-c_{i
\sigma }^{\dagger }c_{j \sigma }\right).
\end{equation}
\begin{equation}
K_x\left(\pmb{r}_i\right)=-\sum _{\sigma } \sum _{j>i} \frac{\left(x_i-x_j\right)^2}{a^2}t_{i j}\left(c_{i \sigma }^{\dagger }c_{j \sigma }+c_{j
\sigma }^{\dagger }c_{i \sigma }\right).
\end{equation}
Here $a=1$ is the lattice constant.
The local conductivity can written in terms of these two operators, 
\begin{equation}
\sigma _{x x}\left(\pmb{r}_i,\omega \right)
=\frac{e^2}{\omega }e^{-i\pmb{q}\pmb{\cdot }\pmb{r}_i}\int _{-\infty }^te^{i \omega  (t-t')}\left\langle \left[J_x^P\left(\pmb{r}_i,t\right),J_x^P(-\pmb{q},t')\right]\right\rangle dt'-\frac{i e^2}{\omega }\left\langle K_x\left(\pmb{r}_i\right)\right\rangle,
\end{equation}
where $\left\langle\cdots\right\rangle$ is the expectation value of the operator. Average over the spatial variable \(\pmb{r}_i\), 

\begin{equation}
\sigma _{x x}(\pmb{q},\omega )=\frac{1}{N}\sum _i \sigma _{x x}\left(\pmb{r}_i,\omega \right)=\frac{e^2}{N \omega }\int _{-\infty }^te^{i \omega  (t-t')}\left\langle \left[J_x^P(\pmb{q},t),J_x^P(-\pmb{q},t')\right]\right\rangle dt'-\frac{ie^2}{\omega }\left\langle K_x\right\rangle,
\end{equation}
where \(\left\langle K_x\right\rangle =\frac{1}{N}\sum _i \left\langle K_x\left(\pmb{r}_i\right)\right\rangle\). The correlation function is only a function of the time difference \(t-t'\), which allows a Fourier transform to frequency domain,
\begin{equation}
\begin{split}
        \sigma _{x x}(\pmb{q},\omega )=\frac{e^2}{i \omega }\left[\frac{i}{N}\int _{-\infty }^{\infty }e^{i \omega  t}\theta (t)\left\langle \left[J_x^P(\pmb{q},t),J_x^P(-\pmb{q},0)\right]\right\rangle dt+\left\langle K_x\right\rangle \right]
        =\frac{e^2}{i \omega }\left[-\Pi _{x x}(\pmb{q},\omega )+\left\langle K_x\right\rangle \right],
\end{split}
\end{equation}
where
\(\Pi _{x x}(\pmb{q},t)=-\frac{i}{N}\theta (t)\left\langle \left[J_x^P(\pmb{q},t),J_x^P(-\pmb{q},0)\right]\right\rangle\)
and 
\(\Pi _{x x}(\pmb{q},\omega )=\int _{-\infty }^{\infty }e^{i \omega  t}\Pi _{x x}(\pmb{q},t)dt\).
The superfluid stiffness $D_s$ is given by,
\begin{equation}
    \frac{D_s}{\pi}=-\left\langle K_x\right\rangle+\Pi _{x x}(q\rightarrow0,\omega=0).
\end{equation}
Here $e$ is set to $1$ and dropped in the final expression.
In the absence of spin-orbit coupling and other spin-flip scattering terms, the dimension of the BdG equation can be reduced from $4N$ to $2N$.
In this case, the BdG transformations are
\begin{equation}
c_{i \uparrow }=\sum _{\tilde{n}}^{'} \left(u_{i \uparrow }^{\tilde{n}_1}\gamma _{\tilde{n}_1}- v_{i \uparrow }^{\tilde{n}_2 *}\gamma _{\tilde{n}_2}^{\dagger
}\right), c_{i \uparrow }^{\dagger }=\sum _{\tilde{n}}^{'} \left(u_{i \uparrow }^{\tilde{n}_1 *}\gamma _{\tilde{n}_1}^{\dagger }- v_{i \uparrow }^{\tilde{n}_2}\gamma
_{\tilde{n}_2}\right).
\end{equation}
\begin{equation}
c_{i \downarrow }=\sum _{\tilde{n}}^{'} \left(u_{i \downarrow }^{\tilde{n}_2}\gamma _{\tilde{n}_2}+ v_{i \downarrow }^{\tilde{n}_1 *}\gamma _{\tilde{n}_1}^{\dagger
}\right), 
c_{i \downarrow }^{\dagger }=\sum _{\tilde{n}}^{'} \left(u_{i \downarrow }^{\tilde{n}_2 *}\gamma _{\tilde{n}_2}^{\dagger }+ v_{i \downarrow
}^{\tilde{n}_1}\gamma _{\tilde{n}_1}\right).
\end{equation}
The prime sign above the summation indicates that only states with positive energy are included. Note that the reduction of Hamiltonian also divides the eigenvalues into two groups and the subscript of \(\tilde{n}_1 \text{ and } \tilde{n}_2\) means
that they correspond to different set of eigenvalues \(E_{\tilde{n}_1}\) and \(E_{\tilde{n}_2}\). Thus there is a set of anti-commutation relations:\(\left\{\gamma _{\tilde{n}_1}^{\dagger },\gamma _{\tilde{m}_1}\right\}=\delta _{\tilde{n}_1\tilde{m}_1},\left\{\gamma _{\tilde{n}_2}^{\dagger },\gamma
_{\tilde{m}_2}\right\}=\delta _{\tilde{n}_2\tilde{m}_2}\),
\(\left\{\gamma _{\tilde{n}_1}^{\dagger },\gamma _{\tilde{m}_2}\right\}=\left\{\gamma _{\tilde{n}_2}^{\dagger },\gamma _{\tilde{m}_1}\right\}=0\).
The kinetic terms can be written in terms of $u$ and $v$ as,
\begin{equation}
\left\langle K_x\right\rangle =-\frac{1}{N}\sum _{n \left(e_n\geq 0\right)} \sum _{i \sigma } \left\{\sum _{j>i} \left(x_i-x_j\right){}^2t_{i j}\left[\left(u_{i \sigma }^{n*}u_{j \sigma }^n+c.c\right)f \left(E_n\right)+\left(v_{j \sigma }^{n*}v_{i \sigma }^n+c.c\right)\left(1-f \left(E_n\right)\right)\right]\right\}.
\end{equation}
Here $f(E)$ is the Fermi function. The current-current correlation function can be written as
\begin{equation}
    \Pi _{x x}(\pmb{q},\omega )=\frac{2}{N}\sum _{\tilde{n}_1,\tilde{m}_1} \left\{\frac{A_{\tilde{n}_1\tilde{m}_1\uparrow }(\pmb{q})\left[A_{\tilde{n}_1\tilde{m}_1\uparrow }^*(\pmb{q})+D_{\tilde{n}_1\tilde{m}_1\downarrow }(-\pmb{q})\right]}{\omega+\left(E_{\tilde{n}_1}-E_{\tilde{m}_1}\right)+\delta  i }\left(f\left( E_{\tilde{n}_1}\right)-f\left( E_{\tilde{m}_1}\right)\right)\right\},
\end{equation}
where
\begin{equation}
A_{n_1 n_2 \sigma}(\pmb{q})=\sum _{i} e^{-i\pmb{q}\cdot \pmb{r}_i}\sum _{j>i} \left(x_j-x_i\right)t_{i j}\left[u_{j \sigma }^{n_1*}u_{i \sigma }^{n_2}-u_{i \sigma }^{n_1*}u_{j \sigma }^{n_2}\right],
\end{equation}
\begin{equation}
D_{n_1 n_2 \sigma}(\pmb{q})=\sum _{i} e^{-i\pmb{q}\cdot \pmb{r}_i}\sum _{j>i} \left(x_j-x_i\right)t_{i j}\left[v_{j \sigma }^{n_1}v_{i \sigma }^{n_2*}-v_{i \sigma }^{n_1}v_{j \sigma }^{n_2*}\right].
\end{equation}
For nearest hopping, $j=i+1$ and the summation $\sum _{j>i} \left(x_j-x_i\right)t_{i j}$ is reduced to $t$. Using the above equations, one can calculate the superfluid stiffness $D_s$ by solving the BdG equation. 

\section{M matrix formulation}

\begin{figure}[hb]
    \centering
    \includegraphics[width = 0.45\linewidth]{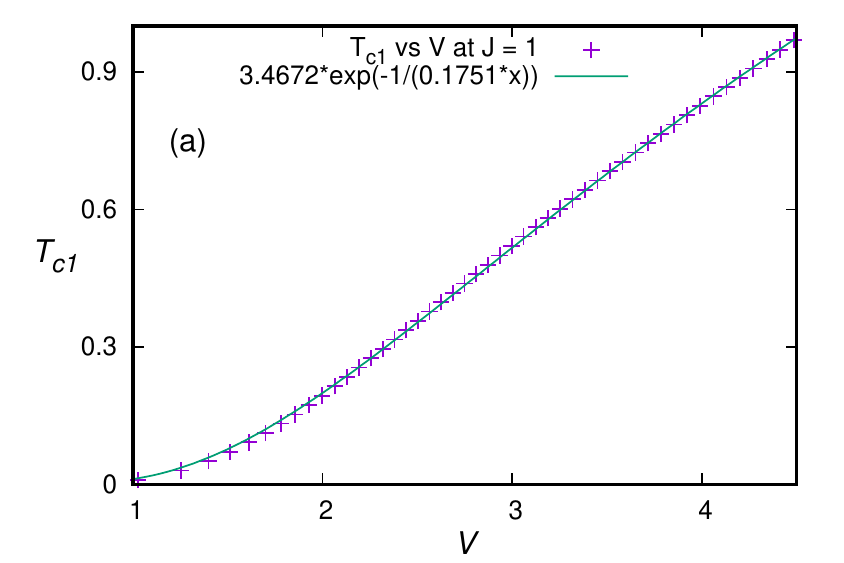}
    \includegraphics[width = 0.45\linewidth]{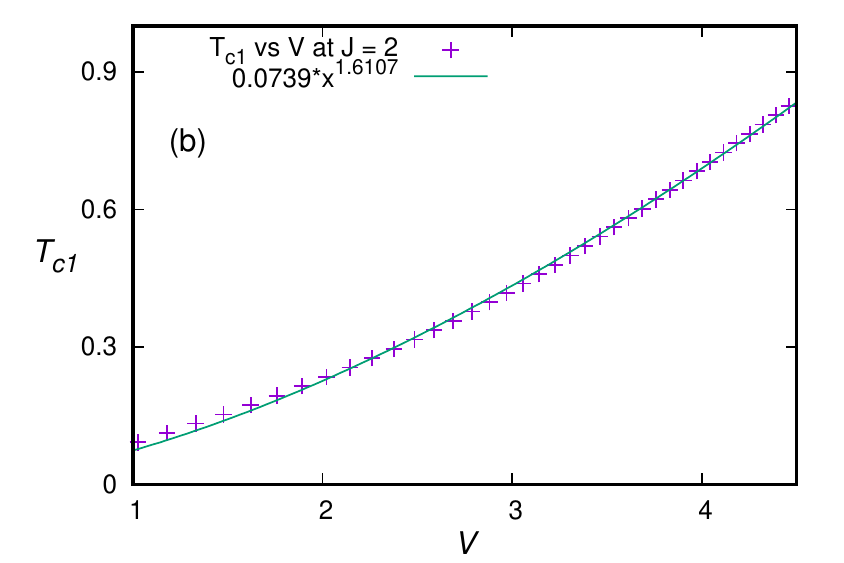}
    \includegraphics[width = 0.45\linewidth]{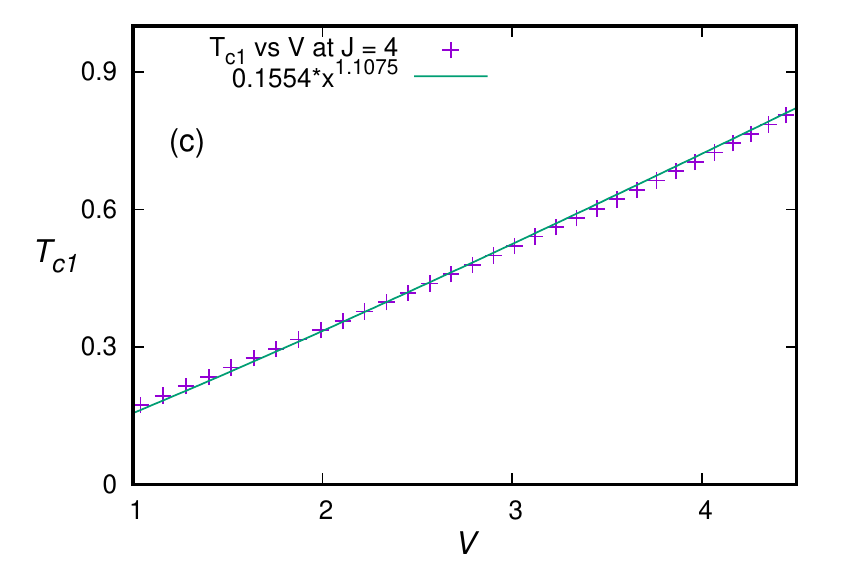}
    \caption{$T_{c}$ vs $V$ calculated using the $M$ matrix formulation for system size $L=987$ at (a) $J=0$, (b) $J=2$, (c) $J=4$.}
    \label{fig:Tc_M_matrix}
\end{figure}

Consider a 1D $s$-wave superconductor with an incommensurate potential described by the Hamiltonian,
\begin{equation}
H=H_{\text{sc}}+H_0,
\end{equation}
\begin{equation}
H_0=-t\sum _{i,j,\sigma} c_{i \sigma }^{\dagger }c_{j\sigma }-J\sum _{i,\sigma } \cos  (2 \pi  i Q)  c_{i \sigma } c_{i \sigma }^{\dagger },
\end{equation}
\begin{equation}
H_{\text{sc}}=-V\sum _i c_{i \uparrow }^{\dagger }c_{i\downarrow }^{\dagger }c_{i\downarrow }c_{i\uparrow },
\end{equation}
where $Q$ is an irrational number and $i$ is the position of site. We set $Q = (\sqrt{5}- 1)/2$. Without superconducting pairing term $H_{\text{sc}}$, the eigenstates are extended for $J/t < 2$ and localized for $ J/t > 2$. The model has a critical point at $J/t = 2$.
The non-interacting Hamiltonian $H_0$ is quadratic and can be diagonalized: $H_0\left|\psi _{\alpha  \sigma }\right\rangle =\epsilon _{\alpha }\left| \psi _{\alpha  \sigma }\right\rangle$. In the weak-coupling $\left| V\right|/{t}\ll 1$, we can rewrite $H$ in this basis and retain pairing interaction only between time reversal partner states,
\begin{equation}
    H'=\sum _{i,\sigma } \epsilon _{\alpha } c_{\alpha  \sigma }^{\dagger } c_{\alpha \sigma } -V\sum_{\alpha,\beta } M_{\alpha \beta }c_{\alpha \uparrow }^{\dagger }c_{\bar{\alpha }\downarrow }^{\dagger }c_{\bar{\beta }\downarrow }c_{\beta \uparrow },
\end{equation}
\begin{equation}
M_{\alpha \beta }=\int \psi _{\alpha}^*  \psi _{\bar{\alpha }}^* \psi _{\beta } \psi _{\bar{\beta }} \, dr=\int \left| \psi _{\alpha }\right|^2 \left| \psi _{\beta }\right|^2 \, dr,
\end{equation}
$\psi _{\bar{\alpha }}$ is the time reversal state of $\psi _{\alpha}$. Using the BCS mean-field approximation, 
$\Delta _{\beta }=V\left\langle c_{\beta \downarrow }c_{\beta \uparrow }\right\rangle$, the self-consistent equation can be therefore written using the $M$ matrix
\begin{equation}
\label{M_matrix_gap_eq}
\Delta _{\alpha }=V \sum _{\beta }  M_{\alpha \beta }\frac{\Delta _{\beta }} {2 E_{\beta }}\tanh{\left( \frac{E_{\beta }}{ 2T}\right) },
\end{equation}
where $E_{\beta }=\sqrt{\Delta _{\beta }^2+\epsilon _{\beta }^2}$. This formulation allows us to study the relation between the critical temperature $T_{c1}$ and the superconducting coupling strength $V$ in the weak coupling regime which is inaccessible for the numerical BdG calculations.

For extended state, all wavefunctions extend over the entire lattice and the amplitude of wavefunctions scales as the inverse square root of the lattice size, \(\left| \psi _{\alpha }\right| \propto 1\left/\sqrt{L}\right.\). Thus, \(M_{\alpha \beta }\) becomes independent of \(\alpha\) and \(\beta\) and scale as \(M_{\alpha \beta }\propto 1\left/L\right.\). The order parameters satisfy \(\Delta _{\alpha }=\Delta _{\beta }=\Delta\), and we obtain
\begin{equation}
\label{extend_m_matrix_1}
    \frac{1}{V}=\sum _{\beta } \frac{1}{L}\frac{1}{2E_{\beta }}\tanh \left(\frac{E_{\beta }}{2T}\right).
\end{equation}
We can transform the discrete summation over states to an integration of energy by introducing the density of state (DOS), \(N(\epsilon )=\frac{1}{L}\sum _{\beta } \delta \left(\epsilon -\epsilon _{\beta }\right)\). In the weak coupling limit, we can approximate \(N(\epsilon )\) by the density of state at Fermi surface \(N_0\) and rewrite equation (\ref{extend_m_matrix_1})\ as
\begin{equation}
    \frac{1}{V}=N_0\int \frac{1}{2E}\tanh \left(\frac{E}{2T}\right) d\epsilon,
\end{equation} 
where \(E=\sqrt{\Delta ^2+\epsilon ^2}\). When the temperature approaches the critical temperature from below \(T\to T_c^-\), the order parameter goes to zero from above \(\Delta \to 0^+\) and \(E\approx \epsilon\). Introducing an integration cutoff \(\hbar \omega _c\gg T_c\), we have
\begin{equation}
    \frac{1}{V N_0}=\int _0^{\hbar \omega _c}\frac{1}{\epsilon }\tanh \left(\frac{\epsilon }{2T_c}\right) d\epsilon
    =\ln \left(\frac{\hbar \omega _c}{2T_c}\right)-\ln  \gamma,
\end{equation}
where \(\gamma\) is a constant number. Therefore, the critical temperature is given by
\begin{equation}
T_c=\frac{\hbar \omega _c}{2\gamma }\exp \left(-\frac{1}{N_0V}\right).
\end{equation}

In the localized state, the wavefunctions are confined in small regions characterized by a localization length \(\xi_l\) and they scale as \(\left|\psi _{\alpha }\right| \propto 1\left/\sqrt{\xi_l }\right.\). The wavefunction has almost no overlap with wavefunctions of other states, as a result, only diagonal terms of \(M_{\alpha \beta }\) are important. Thus the $M$ matrix has the form, \(M_{\alpha \beta }=\delta _{\alpha \beta }/\xi_l\). The gap equation (\ref{M_matrix_gap_eq}) becomes,
\begin{equation}
    \Delta =V\frac{1}{\xi_l }\frac{\Delta }{2E}\tanh \left(\frac{E}{2T}\right),
\end{equation}
where \(E=\sqrt{\Delta ^2+\epsilon ^2}\). When \(\epsilon \gg \Delta\), this equation only has zero solutions \(\Delta =0\). When \(\epsilon \approx
0\), this equation becomes
\begin{equation}
\Delta =V\frac{1}{\xi_l }\frac{\Delta }{2\Delta }\tanh \left(\frac{\Delta }{2T}\right).
\end{equation}
Near the critical temperature, the order parameter \(\Delta\) is small and we can obtain
\begin{equation}
\Delta =\frac{V}{2}\frac{1}{\xi_l }\frac{\Delta }{2T_c}.
\end{equation}
Therefore, the critical temperature is linearly proportional to $V$
\begin{equation}
    T_c=\frac{V}{4\xi_l }.
\end{equation}
In the localized state, $T_c$ increases with $J$ because $\xi_l$ decreases with $J$, which is consistent with the BdG results.
In general cases, the $M$ matrix can be calculated numerically using the wavefunctions obtained from diagonalizing the Hamiltonian. $T_c$ corresponds to the largest eigenvalue of the linearized gap equation. In practice, we find the corresponding $V$ for a given $T_c$. This method gives accurate $T_c$ same as the result obtained from BdG calculation. Results for $T_c$ vs $V$ obtained by $M$ matrix for the AA model in the extended, critical and localized phases are shown in Fig.~ \ref{fig:Tc_M_matrix}.

\section{Additional Results of other 1D incommensurate models}
\subsection{Modulated cosine model}

\begin{figure}[ht]
    \centering
    \includegraphics[width = 0.45\linewidth]{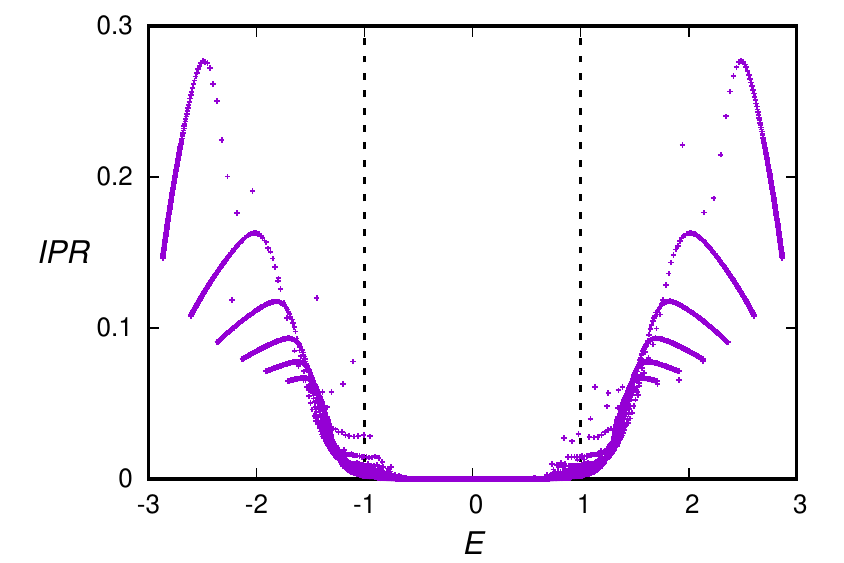}
    \caption{IPR vs $E$ for $J=1$, $\nu=0.7$. The two dot lines at $\mu=\pm 1$ indicate the mobility edges.}
    \label{fig:Modulated_Cos_IPR}
\end{figure}
The modulated cosine model has a modulated incommensurate potential $U_i=J\cos(2\pi Q x_i^\nu)$. The model has two mobility edges at $\mu=\pm J$ and the spectrum is continuous as shown in Fig.~ \ref{fig:Modulated_Cos_IPR}. Here we introduce the inverse participation ratio (IPR) $I_n={\sum_{x_i} |\psi_n(x_i)|^4}{ \left(\sum_{x_i} |\psi_n(x_i)|^2 \right)^{-2}}$, where $\psi_n(x_i)$ is the $n$-th eigenfunction of ${H}_0$. 
$I_n$ is finite for a localized state but vanishes as $1/L^d$ for an extended state. Here 
 $L$ is the linear system size and $d$ the spatial dimension.  The states near the mobility edges are critical which leads to a power-law dependence between $T_c$ and $V$ when $\mu\approx \pm 1$ as shown in the main text.

\subsection{Fibonacci model}
The Fibonacci model has an incommensurate potential: $U_i=U(Qx_i)$ and $U(x)=-J$ for  $m-Q\leqslant x \leqslant m$, $J$ for $m<x<m+1-Q$, where $m$ is an arbitrary integer and $Q=\left(\sqrt{5}-1\right)/2$. The key characteristics of this model is that it is always critical regardless the strength of $J$.
The spectrum exhibits self-similarity, see Fig.~ \ref{fig:Fibonacci_Sepctrum}.

\begin{figure}[h]
    \centering
    \includegraphics[width = 0.45\linewidth]{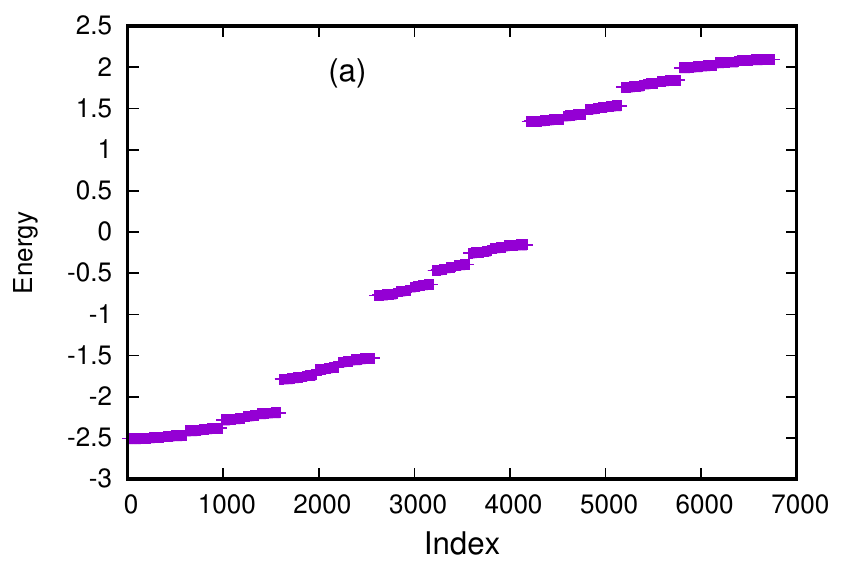}
    \includegraphics[width = 0.45\linewidth]{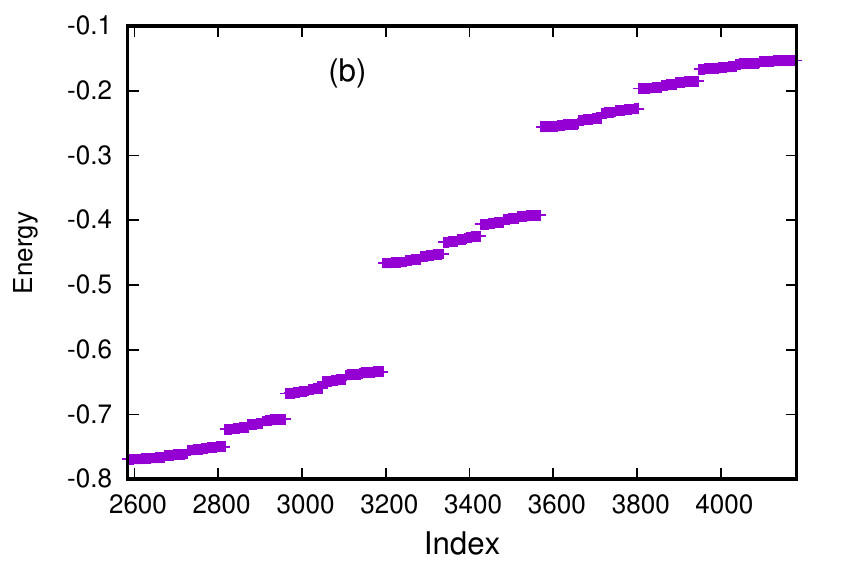}
    \caption{(a) Spectrum of Fibonacci model for system size $L=6765$ and (2) zoomed in spectrum of bands in the center.}
    \label{fig:Fibonacci_Sepctrum}
\end{figure}

\SZL{

\section{Finite size effect}
Here we check the finite size effect. We calculate $T_{c1}$ (temperature when the amplitude of superconducting order parameter vanishes) and $T_{c2}$ (temperature when the superfluid stiffness vanishes) for different system sizes, and the results are shown in Fig.~\ref{fig:Finite_size}. It can be seen that $T_{c1}$ and $T_{c2}$ converge to a fixed value very quickly as one increases $L$. Therefore, the system size with $L=233$ we used in the main text has negligible finite size effect for the $V$ we used. When $V$ is reduced, larger system size is required because the superconducting coherence length increases when $V$ is reduced. 

\begin{figure}
    \centering
    \includegraphics{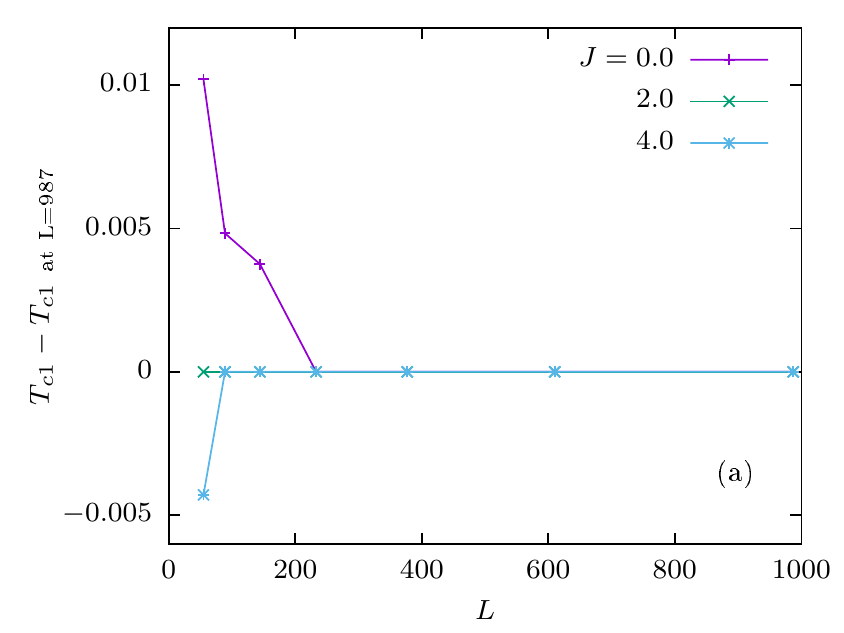}
    \includegraphics{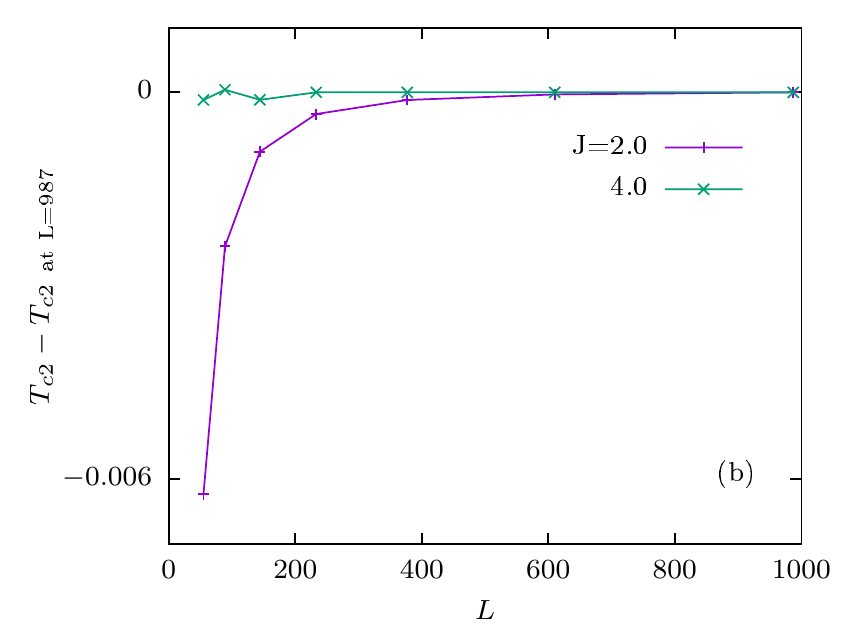}
    \caption{ (a) $T_{c1}$  and (b) $T_{c2}$ as a function of system size $L$. Here $V=1$.}
    \label{fig:Finite_size}
\end{figure}

\section{Density of state in the normal state}
Superconducting transition temperature $T_c$ depends on the normal state density of state (DOS) at the Fermi energy. Here we calculate the normal state DOS in the presence of an incommensurate potential using the Aubry-Andr\'{e} model in Eq.~\eqref{AA_Hamiltonian:NI}. As displayed in Fig.~\ref{fig:normal_dos}, the DOS at Fermi energy is increased in the presence of an incommensurate potential. However, this increase of DOS cannot explain the power law dependence of $T_c$ on the pairing interaction.

\begin{figure}
    \centering
    \includegraphics{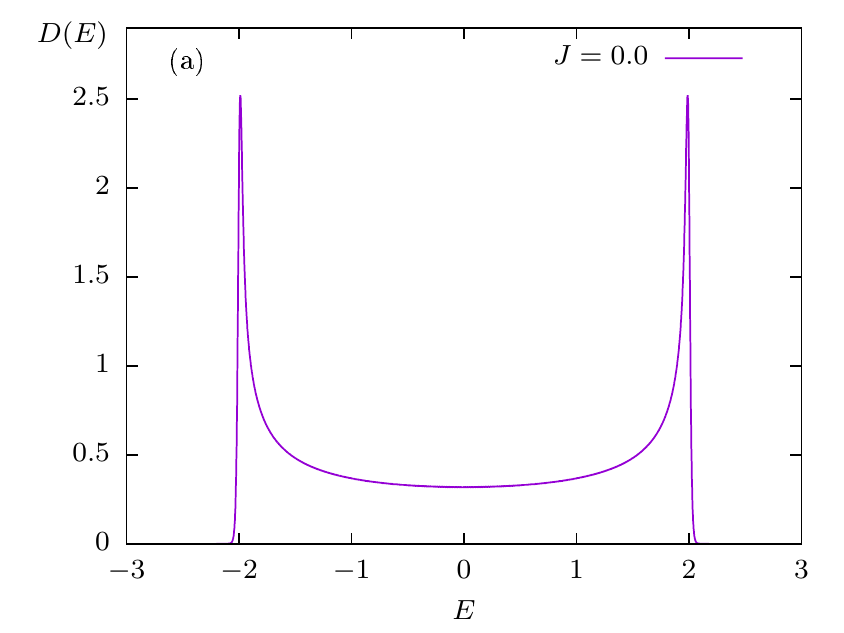}
    \includegraphics{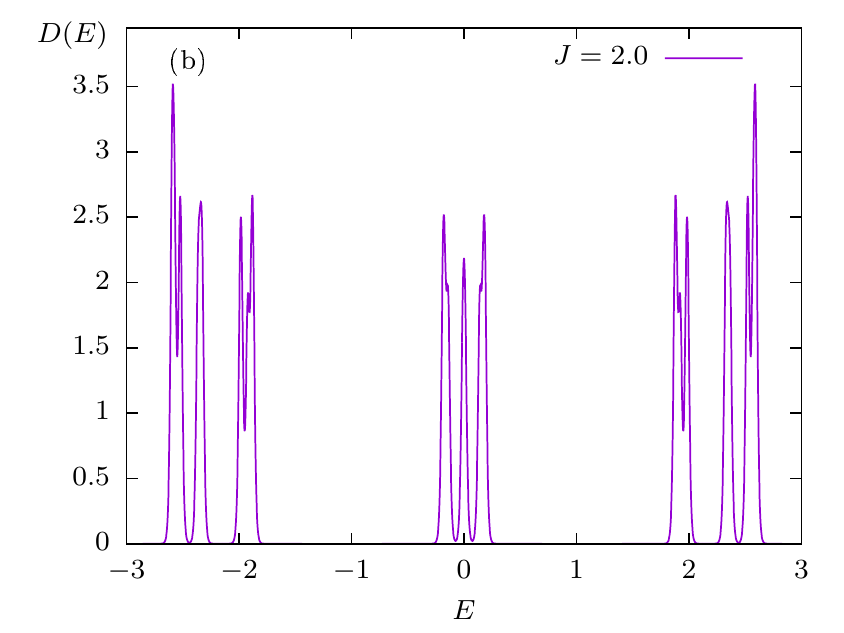}
    \caption{Normal state DOS at $J=0.0$ (a) and $J=2.0$ (b).}
    \label{fig:normal_dos}
\end{figure}

\section{Comparison to the random disordered model}
The random disorders have stronger localization effect compared to the incommensurate potential. For example, in 1D and 2D systems without spin orbit coupling, arbitrarily weak random disorders cause localization of electrons. In this work, we use an incommensurate potential, which allows us to access the localization transition at a nonzero incommensurate potential strength. The choice of the incommensurate potential is also motivated by the recent exciting experimental discovery of superconductivity in graphene Moir\'{e} superlattices.
To compare the results with an incommensurate potential and random disorders, we perform additional BdG calculations of 1D $s$-wave superconductor with an random potential. 


The Hamiltonian is given by
\begin{equation}
H=H_0+H_{\text{sc}},
\end{equation}
\begin{equation}
H_0=-t\sum _{i,j,\sigma} c_{i \sigma }^{\dagger }c_{j\sigma }-\sum _{i,\sigma } J_i c_{i \sigma }^{\dagger } c_{i \sigma } ,
\end{equation}
\begin{equation}
H_{\text{sc}}=-V\sum _i c_{i \uparrow }^{\dagger }c_{i\downarrow }^{\dagger }c_{i\downarrow }c_{i\uparrow }, 
\end{equation}
where $J_i$ is a random potential at the $i$-th site that is uniformly distributed in $[-J,+J]$.}
\begin{figure}[h]
    \centering
    \includegraphics{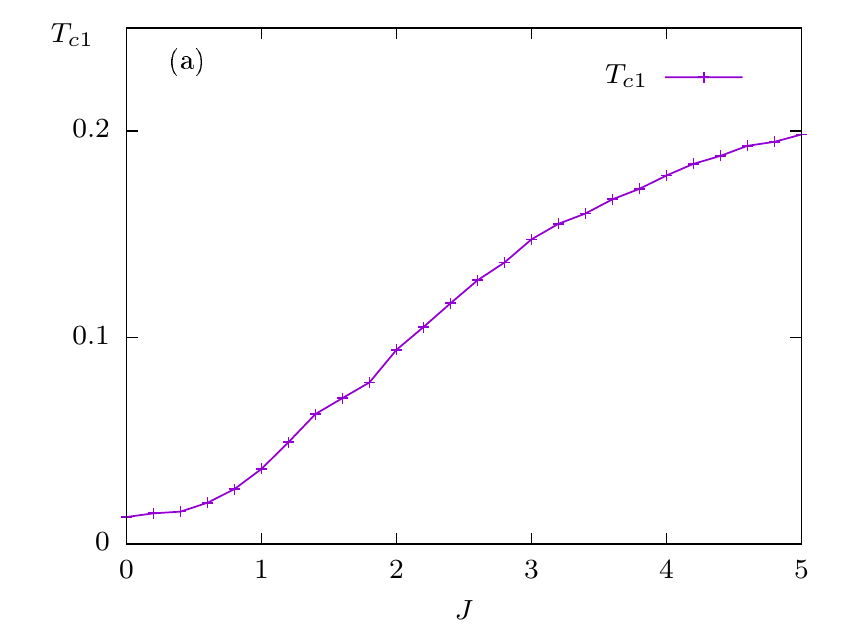}
    \includegraphics{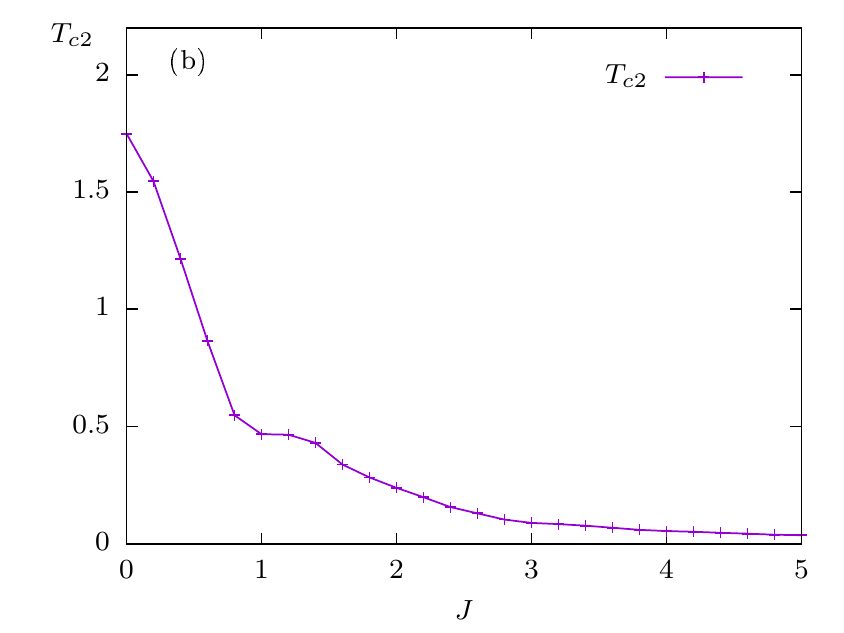}
    \caption{$T_{c1}$ and $T_{c2}$ vs $J$ for random disordered system with $L=233$ and $V=1$. The results are obtained by averaging over 5 independent random disorder configurations.}
    \label{fig:Tc_random_potential}
\end{figure}
The system size is $L=233$ with $V =1$, and the chemical potential is tuned to keep the system half-filled. Note that in one dimension, the localization transition occurs at $J=0$. In Fig.~\ref{fig:Tc_random_potential},  we show $T_{c1}$ and $T_{c2}$ as functions of $J$. $T_{c1}$ increases while $T_{c2}$ decreases with $J$, which is qualitatively similar to that for an incommensurate potentials in the localized region [see Fig.~\ref{fig:AA_BdG_results} (a)].

\end{appendix}
\end{widetext}
\bibliography{References.bib}

\providecommand{\noopsort}[1]{}\providecommand{\singleletter}[1]{#1}%
\begin{thebibliography}{40}%
\makeatletter
\providecommand \@ifxundefined [1]{%
 \@ifx{#1\undefined}
}%
\providecommand \@ifnum [1]{%
 \ifnum #1\expandafter \@firstoftwo
 \else \expandafter \@secondoftwo
 \fi
}%
\providecommand \@ifx [1]{%
 \ifx #1\expandafter \@firstoftwo
 \else \expandafter \@secondoftwo
 \fi
}%
\providecommand \natexlab [1]{#1}%
\providecommand \enquote  [1]{``#1''}%
\providecommand \bibnamefont  [1]{#1}%
\providecommand \bibfnamefont [1]{#1}%
\providecommand \citenamefont [1]{#1}%
\providecommand \href@noop [0]{\@secondoftwo}%
\providecommand \href [0]{\begingroup \@sanitize@url \@href}%
\providecommand \@href[1]{\@@startlink{#1}\@@href}%
\providecommand \@@href[1]{\endgroup#1\@@endlink}%
\providecommand \@sanitize@url [0]{\catcode `\\12\catcode `\$12\catcode
  `\&12\catcode `\#12\catcode `\^12\catcode `\_12\catcode `\%12\relax}%
\providecommand \@@startlink[1]{}%
\providecommand \@@endlink[0]{}%
\providecommand \url  [0]{\begingroup\@sanitize@url \@url }%
\providecommand \@url [1]{\endgroup\@href {#1}{\urlprefix }}%
\providecommand \urlprefix  [0]{URL }%
\providecommand \Eprint [0]{\href }%
\providecommand \doibase [0]{http://dx.doi.org/}%
\providecommand \selectlanguage [0]{\@gobble}%
\providecommand \bibinfo  [0]{\@secondoftwo}%
\providecommand \bibfield  [0]{\@secondoftwo}%
\providecommand \translation [1]{[#1]}%
\providecommand \BibitemOpen [0]{}%
\providecommand \bibitemStop [0]{}%
\providecommand \bibitemNoStop [0]{.\EOS\space}%
\providecommand \EOS [0]{\spacefactor3000\relax}%
\providecommand \BibitemShut  [1]{\csname bibitem#1\endcsname}%
\let\auto@bib@innerbib\@empty
\bibitem [{\citenamefont {Siebesma}\ and\ \citenamefont
  {Pietronero}(1987)}]{Siebesma_1987_Multifractal_Properties_of_Wave_Functions_for_One-Dimensional_Systems_with_an_Incommensurate_Potential}%
  \BibitemOpen
  \bibfield  {author} {\bibinfo {author} {\bibfnamefont {A.~P.}\ \bibnamefont
  {Siebesma}}\ and\ \bibinfo {author} {\bibfnamefont {L.}~\bibnamefont
  {Pietronero}},\ }\href {\doibase 10.1209/0295-5075/4/5/014} {\bibfield
  {journal} {\bibinfo  {journal} {Europhysics Letters ({EPL})}\ }\textbf
  {\bibinfo {volume} {4}},\ \bibinfo {pages} {597} (\bibinfo {year}
  {1987})}\BibitemShut {NoStop}%
\bibitem [{\citenamefont {Aubry}\ and\ \citenamefont
  {Andr{\'e}}(1980)}]{Aubry_Analyticity_1980}%
  \BibitemOpen
  \bibfield  {author} {\bibinfo {author} {\bibfnamefont {S.}~\bibnamefont
  {Aubry}}\ and\ \bibinfo {author} {\bibfnamefont {G.}~\bibnamefont
  {Andr{\'e}}},\ }\href@noop {} {\bibfield  {journal} {\bibinfo  {journal}
  {Ann. Israel Phys. Soc}\ }\textbf {\bibinfo {volume} {3}},\ \bibinfo {pages}
  {18} (\bibinfo {year} {1980})}\BibitemShut {NoStop}%
\bibitem [{\citenamefont {Devakul}\ and\ \citenamefont
  {Huse}(2017)}]{PhysRevB.96.214201}%
  \BibitemOpen
  \bibfield  {author} {\bibinfo {author} {\bibfnamefont {T.}~\bibnamefont
  {Devakul}}\ and\ \bibinfo {author} {\bibfnamefont {D.~A.}\ \bibnamefont
  {Huse}},\ }\href {\doibase 10.1103/PhysRevB.96.214201} {\bibfield  {journal}
  {\bibinfo  {journal} {Phys. Rev. B}\ }\textbf {\bibinfo {volume} {96}},\
  \bibinfo {pages} {214201} (\bibinfo {year} {2017})}\BibitemShut {NoStop}%
\bibitem [{\citenamefont {Su}\ and\ \citenamefont
  {Lin}(2018)}]{PhysRevB.98.235116}%
  \BibitemOpen
  \bibfield  {author} {\bibinfo {author} {\bibfnamefont {Y.}~\bibnamefont
  {Su}}\ and\ \bibinfo {author} {\bibfnamefont {S.-Z.}\ \bibnamefont {Lin}},\
  }\href {\doibase 10.1103/PhysRevB.98.235116} {\bibfield  {journal} {\bibinfo
  {journal} {Phys. Rev. B}\ }\textbf {\bibinfo {volume} {98}},\ \bibinfo
  {pages} {235116} (\bibinfo {year} {2018})}\BibitemShut {NoStop}%
\bibitem [{\citenamefont {Arai}\ \emph {et~al.}(1988)\citenamefont {Arai},
  \citenamefont {Tokihiro}, \citenamefont {Fujiwara},\ and\ \citenamefont
  {Kohmoto}}]{PhysRevB.38.1621}%
  \BibitemOpen
  \bibfield  {author} {\bibinfo {author} {\bibfnamefont {M.}~\bibnamefont
  {Arai}}, \bibinfo {author} {\bibfnamefont {T.}~\bibnamefont {Tokihiro}},
  \bibinfo {author} {\bibfnamefont {T.}~\bibnamefont {Fujiwara}}, \ and\
  \bibinfo {author} {\bibfnamefont {M.}~\bibnamefont {Kohmoto}},\ }\href
  {\doibase 10.1103/PhysRevB.38.1621} {\bibfield  {journal} {\bibinfo
  {journal} {Phys. Rev. B}\ }\textbf {\bibinfo {volume} {38}},\ \bibinfo
  {pages} {1621} (\bibinfo {year} {1988})}\BibitemShut {NoStop}%
\bibitem [{\citenamefont {Kraj}\ and\ \citenamefont
  {Fujiwara}(1988)}]{PhysRevB.38.12903}%
  \BibitemOpen
  \bibfield  {author} {\bibinfo {author} {\bibfnamefont {M.}~\bibnamefont
  {Kraj}}\ and\ \bibinfo {author} {\bibfnamefont {T.}~\bibnamefont
  {Fujiwara}},\ }\href {\doibase 10.1103/PhysRevB.38.12903} {\bibfield
  {journal} {\bibinfo  {journal} {Phys. Rev. B}\ }\textbf {\bibinfo {volume}
  {38}},\ \bibinfo {pages} {12903} (\bibinfo {year} {1988})}\BibitemShut
  {NoStop}%
\bibitem [{\citenamefont {Tokihiro}\ \emph {et~al.}(1988)\citenamefont
  {Tokihiro}, \citenamefont {Fujiwara},\ and\ \citenamefont
  {Arai}}]{PhysRevB.38.5981}%
  \BibitemOpen
  \bibfield  {author} {\bibinfo {author} {\bibfnamefont {T.}~\bibnamefont
  {Tokihiro}}, \bibinfo {author} {\bibfnamefont {T.}~\bibnamefont {Fujiwara}},
  \ and\ \bibinfo {author} {\bibfnamefont {M.}~\bibnamefont {Arai}},\ }\href
  {\doibase 10.1103/PhysRevB.38.5981} {\bibfield  {journal} {\bibinfo
  {journal} {Phys. Rev. B}\ }\textbf {\bibinfo {volume} {38}},\ \bibinfo
  {pages} {5981} (\bibinfo {year} {1988})}\BibitemShut {NoStop}%
\bibitem [{\citenamefont {Kamiya}\ \emph {et~al.}(2018)\citenamefont {Kamiya},
  \citenamefont {Takeuchi}, \citenamefont {Kabeya}, \citenamefont {Wada},
  \citenamefont {Ishimasa}, \citenamefont {Ochiai}, \citenamefont {Deguchi},
  \citenamefont {Imura},\ and\ \citenamefont {Sato}}]{kamiya_discovery_2018}%
  \BibitemOpen
  \bibfield  {author} {\bibinfo {author} {\bibfnamefont {K.}~\bibnamefont
  {Kamiya}}, \bibinfo {author} {\bibfnamefont {T.}~\bibnamefont {Takeuchi}},
  \bibinfo {author} {\bibfnamefont {N.}~\bibnamefont {Kabeya}}, \bibinfo
  {author} {\bibfnamefont {N.}~\bibnamefont {Wada}}, \bibinfo {author}
  {\bibfnamefont {T.}~\bibnamefont {Ishimasa}}, \bibinfo {author}
  {\bibfnamefont {A.}~\bibnamefont {Ochiai}}, \bibinfo {author} {\bibfnamefont
  {K.}~\bibnamefont {Deguchi}}, \bibinfo {author} {\bibfnamefont
  {K.}~\bibnamefont {Imura}}, \ and\ \bibinfo {author} {\bibfnamefont {N.~K.}\
  \bibnamefont {Sato}},\ }\href {\doibase 10.1038/s41467-017-02667-x}
  {\bibfield  {journal} {\bibinfo  {journal} {Nature Communications}\ }\textbf
  {\bibinfo {volume} {9}},\ \bibinfo {pages} {1} (\bibinfo {year} {2018})},\
  \bibinfo {note} {number: 1 Publisher: Nature Publishing Group}\BibitemShut
  {NoStop}%
\bibitem [{\citenamefont {Deguchi}\ \emph {et~al.}(2012)\citenamefont
  {Deguchi}, \citenamefont {Matsukawa}, \citenamefont {Sato}, \citenamefont
  {Hattori}, \citenamefont {Ishida}, \citenamefont {Takakura},\ and\
  \citenamefont {Ishimasa}}]{deguchi_quantum_2012}%
  \BibitemOpen
  \bibfield  {author} {\bibinfo {author} {\bibfnamefont {K.}~\bibnamefont
  {Deguchi}}, \bibinfo {author} {\bibfnamefont {S.}~\bibnamefont {Matsukawa}},
  \bibinfo {author} {\bibfnamefont {N.~K.}\ \bibnamefont {Sato}}, \bibinfo
  {author} {\bibfnamefont {T.}~\bibnamefont {Hattori}}, \bibinfo {author}
  {\bibfnamefont {K.}~\bibnamefont {Ishida}}, \bibinfo {author} {\bibfnamefont
  {H.}~\bibnamefont {Takakura}}, \ and\ \bibinfo {author} {\bibfnamefont
  {T.}~\bibnamefont {Ishimasa}},\ }\href {\doibase 10.1038/nmat3432} {\bibfield
   {journal} {\bibinfo  {journal} {Nature Materials}\ }\textbf {\bibinfo
  {volume} {11}},\ \bibinfo {pages} {1013} (\bibinfo {year} {2012})},\ \bibinfo
  {note} {number: 12 Publisher: Nature Publishing Group}\BibitemShut {NoStop}%
\bibitem [{\citenamefont {Cao}\ \emph {et~al.}(2018{\natexlab{a}})\citenamefont
  {Cao}, \citenamefont {Fatemi}, \citenamefont {Demir}, \citenamefont {Fang},
  \citenamefont {Tomarken}, \citenamefont {Luo}, \citenamefont
  {Sanchez-Yamagishi}, \citenamefont {Watanabe}, \citenamefont {Taniguchi},
  \citenamefont {Kaxiras}, \citenamefont {Ashoori},\ and\ \citenamefont
  {Jarillo-Herrero}}]{cao_correlated_2018}%
  \BibitemOpen
  \bibfield  {author} {\bibinfo {author} {\bibfnamefont {Y.}~\bibnamefont
  {Cao}}, \bibinfo {author} {\bibfnamefont {V.}~\bibnamefont {Fatemi}},
  \bibinfo {author} {\bibfnamefont {A.}~\bibnamefont {Demir}}, \bibinfo
  {author} {\bibfnamefont {S.}~\bibnamefont {Fang}}, \bibinfo {author}
  {\bibfnamefont {S.~L.}\ \bibnamefont {Tomarken}}, \bibinfo {author}
  {\bibfnamefont {J.~Y.}\ \bibnamefont {Luo}}, \bibinfo {author} {\bibfnamefont
  {J.~D.}\ \bibnamefont {Sanchez-Yamagishi}}, \bibinfo {author} {\bibfnamefont
  {K.}~\bibnamefont {Watanabe}}, \bibinfo {author} {\bibfnamefont
  {T.}~\bibnamefont {Taniguchi}}, \bibinfo {author} {\bibfnamefont
  {E.}~\bibnamefont {Kaxiras}}, \bibinfo {author} {\bibfnamefont {R.~C.}\
  \bibnamefont {Ashoori}}, \ and\ \bibinfo {author} {\bibfnamefont
  {P.}~\bibnamefont {Jarillo-Herrero}},\ }\href {\doibase 10.1038/nature26154}
  {\bibfield  {journal} {\bibinfo  {journal} {Nature}\ }\textbf {\bibinfo
  {volume} {556}},\ \bibinfo {pages} {80} (\bibinfo {year}
  {2018}{\natexlab{a}})}\BibitemShut {NoStop}%
\bibitem [{\citenamefont {Cao}\ \emph {et~al.}(2018{\natexlab{b}})\citenamefont
  {Cao}, \citenamefont {Fatemi}, \citenamefont {Fang}, \citenamefont
  {Watanabe}, \citenamefont {Taniguchi}, \citenamefont {Kaxiras},\ and\
  \citenamefont {Jarillo-Herrero}}]{cao_unconventional_2018}%
  \BibitemOpen
  \bibfield  {author} {\bibinfo {author} {\bibfnamefont {Y.}~\bibnamefont
  {Cao}}, \bibinfo {author} {\bibfnamefont {V.}~\bibnamefont {Fatemi}},
  \bibinfo {author} {\bibfnamefont {S.}~\bibnamefont {Fang}}, \bibinfo {author}
  {\bibfnamefont {K.}~\bibnamefont {Watanabe}}, \bibinfo {author}
  {\bibfnamefont {T.}~\bibnamefont {Taniguchi}}, \bibinfo {author}
  {\bibfnamefont {E.}~\bibnamefont {Kaxiras}}, \ and\ \bibinfo {author}
  {\bibfnamefont {P.}~\bibnamefont {Jarillo-Herrero}},\ }\href {\doibase
  10.1038/nature26160} {\bibfield  {journal} {\bibinfo  {journal} {Nature}\
  }\textbf {\bibinfo {volume} {556}},\ \bibinfo {pages} {43} (\bibinfo {year}
  {2018}{\natexlab{b}})}\BibitemShut {NoStop}%
\bibitem [{\citenamefont {Yankowitz}\ \emph {et~al.}(2019)\citenamefont
  {Yankowitz}, \citenamefont {Chen}, \citenamefont {Polshyn}, \citenamefont
  {Zhang}, \citenamefont {Watanabe}, \citenamefont {Taniguchi}, \citenamefont
  {Graf}, \citenamefont {Young},\ and\ \citenamefont {Dean}}]{Yankowitz1059}%
  \BibitemOpen
  \bibfield  {author} {\bibinfo {author} {\bibfnamefont {M.}~\bibnamefont
  {Yankowitz}}, \bibinfo {author} {\bibfnamefont {S.}~\bibnamefont {Chen}},
  \bibinfo {author} {\bibfnamefont {H.}~\bibnamefont {Polshyn}}, \bibinfo
  {author} {\bibfnamefont {Y.}~\bibnamefont {Zhang}}, \bibinfo {author}
  {\bibfnamefont {K.}~\bibnamefont {Watanabe}}, \bibinfo {author}
  {\bibfnamefont {T.}~\bibnamefont {Taniguchi}}, \bibinfo {author}
  {\bibfnamefont {D.}~\bibnamefont {Graf}}, \bibinfo {author} {\bibfnamefont
  {A.~F.}\ \bibnamefont {Young}}, \ and\ \bibinfo {author} {\bibfnamefont
  {C.~R.}\ \bibnamefont {Dean}},\ }\href {\doibase 10.1126/science.aav1910}
  {\bibfield  {journal} {\bibinfo  {journal} {Science}\ }\textbf {\bibinfo
  {volume} {363}},\ \bibinfo {pages} {1059} (\bibinfo {year}
  {2019})}\BibitemShut {NoStop}%
\bibitem [{\citenamefont {Lu}\ \emph {et~al.}(2019)\citenamefont {Lu},
  \citenamefont {Stepanov}, \citenamefont {Yang}, \citenamefont {Xie},
  \citenamefont {Aamir}, \citenamefont {Das}, \citenamefont {Urgell},
  \citenamefont {Watanabe}, \citenamefont {Taniguchi}, \citenamefont {Zhang}
  \emph {et~al.}}]{lu2019superconductors}%
  \BibitemOpen
  \bibfield  {author} {\bibinfo {author} {\bibfnamefont {X.}~\bibnamefont
  {Lu}}, \bibinfo {author} {\bibfnamefont {P.}~\bibnamefont {Stepanov}},
  \bibinfo {author} {\bibfnamefont {W.}~\bibnamefont {Yang}}, \bibinfo {author}
  {\bibfnamefont {M.}~\bibnamefont {Xie}}, \bibinfo {author} {\bibfnamefont
  {M.~A.}\ \bibnamefont {Aamir}}, \bibinfo {author} {\bibfnamefont
  {I.}~\bibnamefont {Das}}, \bibinfo {author} {\bibfnamefont {C.}~\bibnamefont
  {Urgell}}, \bibinfo {author} {\bibfnamefont {K.}~\bibnamefont {Watanabe}},
  \bibinfo {author} {\bibfnamefont {T.}~\bibnamefont {Taniguchi}}, \bibinfo
  {author} {\bibfnamefont {G.}~\bibnamefont {Zhang}},  \emph {et~al.},\ }\href
  {\doibase 10.1038/s41586-019-1695-0} {\bibfield  {journal} {\bibinfo
  {journal} {Nature}\ }\textbf {\bibinfo {volume} {574}},\ \bibinfo {pages}
  {653} (\bibinfo {year} {2019})}\BibitemShut {NoStop}%
\bibitem [{\citenamefont {Lopes~dos Santos}\ \emph {et~al.}(2012)\citenamefont
  {Lopes~dos Santos}, \citenamefont {Peres},\ and\ \citenamefont
  {Castro~Neto}}]{PhysRevB.86.155449}%
  \BibitemOpen
  \bibfield  {author} {\bibinfo {author} {\bibfnamefont {J.~M.~B.}\
  \bibnamefont {Lopes~dos Santos}}, \bibinfo {author} {\bibfnamefont
  {N.~M.~R.}\ \bibnamefont {Peres}}, \ and\ \bibinfo {author} {\bibfnamefont
  {A.~H.}\ \bibnamefont {Castro~Neto}},\ }\href {\doibase
  10.1103/PhysRevB.86.155449} {\bibfield  {journal} {\bibinfo  {journal} {Phys.
  Rev. B}\ }\textbf {\bibinfo {volume} {86}},\ \bibinfo {pages} {155449}
  (\bibinfo {year} {2012})}\BibitemShut {NoStop}%
\bibitem [{\citenamefont {Bistritzer}\ and\ \citenamefont
  {MacDonald}(2011)}]{bistritzer_moire_2011}%
  \BibitemOpen
  \bibfield  {author} {\bibinfo {author} {\bibfnamefont {R.}~\bibnamefont
  {Bistritzer}}\ and\ \bibinfo {author} {\bibfnamefont {A.~H.}\ \bibnamefont
  {MacDonald}},\ }\href {\doibase 10.1073/pnas.1108174108} {\bibfield
  {journal} {\bibinfo  {journal} {PNAS}\ }\textbf {\bibinfo {volume} {108}},\
  \bibinfo {pages} {12233} (\bibinfo {year} {2011})}\BibitemShut {NoStop}%
\bibitem [{\citenamefont {Huang}\ and\ \citenamefont
  {Liu}(2019)}]{PhysRevB.100.144202}%
  \BibitemOpen
  \bibfield  {author} {\bibinfo {author} {\bibfnamefont {B.}~\bibnamefont
  {Huang}}\ and\ \bibinfo {author} {\bibfnamefont {W.~V.}\ \bibnamefont
  {Liu}},\ }\href {\doibase 10.1103/PhysRevB.100.144202} {\bibfield  {journal}
  {\bibinfo  {journal} {Phys. Rev. B}\ }\textbf {\bibinfo {volume} {100}},\
  \bibinfo {pages} {144202} (\bibinfo {year} {2019})}\BibitemShut {NoStop}%
\bibitem [{\citenamefont {Hiramoto}\ and\ \citenamefont
  {Kohmoto}(1992)}]{HIRAMOTO_A_SCALING_APPROACH}%
  \BibitemOpen
  \bibfield  {author} {\bibinfo {author} {\bibfnamefont {H.}~\bibnamefont
  {Hiramoto}}\ and\ \bibinfo {author} {\bibfnamefont {M.}~\bibnamefont
  {Kohmoto}},\ }\href {\doibase 10.1142/S0217979292000153} {\bibfield
  {journal} {\bibinfo  {journal} {International Journal of Modern Physics B}\
  }\textbf {\bibinfo {volume} {06}},\ \bibinfo {pages} {281} (\bibinfo {year}
  {1992})},\ \Eprint
  {http://arxiv.org/abs/https://doi.org/10.1142/S0217979292000153}
  {https://doi.org/10.1142/S0217979292000153} \BibitemShut {NoStop}%
\bibitem [{\citenamefont {Aubry}\ and\ \citenamefont
  {André}(1980)}]{Aubry_Andre_original}%
  \BibitemOpen
  \bibfield  {author} {\bibinfo {author} {\bibfnamefont {S.}~\bibnamefont
  {Aubry}}\ and\ \bibinfo {author} {\bibfnamefont {G.}~\bibnamefont {André}},\
  }\href@noop {} {\bibfield  {journal} {\bibinfo  {journal} {Proceedings, VIII
  International Colloquium on Group-Theoretical Methods in Physics}\ }\textbf
  {\bibinfo {volume} {3}} (\bibinfo {year} {1980})}\BibitemShut {NoStop}%
\bibitem [{\citenamefont {Zhu}(2016)}]{Zhu_2016}%
  \BibitemOpen
  \bibfield  {author} {\bibinfo {author} {\bibfnamefont {J.-X.}\ \bibnamefont
  {Zhu}},\ }\href {\doibase 10.1007/978-3-319-31314-6} {\emph {\bibinfo {title}
  {Bogoliubov-de Gennes Method and Its Applications}}}\ (\bibinfo  {publisher}
  {Springer International Publishing},\ \bibinfo {year} {2016})\BibitemShut
  {NoStop}%
\bibitem [{\citenamefont {Scalapino}\ \emph {et~al.}(1993)\citenamefont
  {Scalapino}, \citenamefont {White},\ and\ \citenamefont
  {Zhang}}]{PhysRevB.47.7995}%
  \BibitemOpen
  \bibfield  {author} {\bibinfo {author} {\bibfnamefont {D.~J.}\ \bibnamefont
  {Scalapino}}, \bibinfo {author} {\bibfnamefont {S.~R.}\ \bibnamefont
  {White}}, \ and\ \bibinfo {author} {\bibfnamefont {S.}~\bibnamefont
  {Zhang}},\ }\href {\doibase 10.1103/PhysRevB.47.7995} {\bibfield  {journal}
  {\bibinfo  {journal} {Phys. Rev. B}\ }\textbf {\bibinfo {volume} {47}},\
  \bibinfo {pages} {7995} (\bibinfo {year} {1993})}\BibitemShut {NoStop}%
\bibitem [{\citenamefont {Ghosal}\ \emph {et~al.}(2001)\citenamefont {Ghosal},
  \citenamefont {Randeria},\ and\ \citenamefont {Trivedi}}]{Nandini2001}%
  \BibitemOpen
  \bibfield  {author} {\bibinfo {author} {\bibfnamefont {A.}~\bibnamefont
  {Ghosal}}, \bibinfo {author} {\bibfnamefont {M.}~\bibnamefont {Randeria}}, \
  and\ \bibinfo {author} {\bibfnamefont {N.}~\bibnamefont {Trivedi}},\ }\href
  {\doibase 10.1103/PhysRevB.65.014501} {\bibfield  {journal} {\bibinfo
  {journal} {Phys. Rev. B}\ }\textbf {\bibinfo {volume} {65}},\ \bibinfo
  {pages} {014501} (\bibinfo {year} {2001})}\BibitemShut {NoStop}%
\bibitem [{\citenamefont {Ma}\ and\ \citenamefont
  {Lee}(1985)}]{PhysRevB.32.5658}%
  \BibitemOpen
  \bibfield  {author} {\bibinfo {author} {\bibfnamefont {M.}~\bibnamefont
  {Ma}}\ and\ \bibinfo {author} {\bibfnamefont {P.~A.}\ \bibnamefont {Lee}},\
  }\href {\doibase 10.1103/PhysRevB.32.5658} {\bibfield  {journal} {\bibinfo
  {journal} {Phys. Rev. B}\ }\textbf {\bibinfo {volume} {32}},\ \bibinfo
  {pages} {5658} (\bibinfo {year} {1985})}\BibitemShut {NoStop}%
\bibitem [{\citenamefont
  {Kohmoto}(1983)}]{PhysRevLett.51.1198_Fibonacci_Model}%
  \BibitemOpen
  \bibfield  {author} {\bibinfo {author} {\bibfnamefont {M.}~\bibnamefont
  {Kohmoto}},\ }\href {\doibase 10.1103/PhysRevLett.51.1198} {\bibfield
  {journal} {\bibinfo  {journal} {Phys. Rev. Lett.}\ }\textbf {\bibinfo
  {volume} {51}},\ \bibinfo {pages} {1198} (\bibinfo {year}
  {1983})}\BibitemShut {NoStop}%
\bibitem [{\citenamefont {Anderson}(1959)}]{ANDERSON195926}%
  \BibitemOpen
  \bibfield  {author} {\bibinfo {author} {\bibfnamefont {P.}~\bibnamefont
  {Anderson}},\ }\href {\doibase https://doi.org/10.1016/0022-3697(59)90036-8}
  {\bibfield  {journal} {\bibinfo  {journal} {Journal of Physics and Chemistry
  of Solids}\ }\textbf {\bibinfo {volume} {11}},\ \bibinfo {pages} {26 }
  (\bibinfo {year} {1959})}\BibitemShut {NoStop}%
\bibitem [{\citenamefont {Tang}\ and\ \citenamefont
  {Kohmoto}(1986)}]{PhysRevB.34.2041}%
  \BibitemOpen
  \bibfield  {author} {\bibinfo {author} {\bibfnamefont {C.}~\bibnamefont
  {Tang}}\ and\ \bibinfo {author} {\bibfnamefont {M.}~\bibnamefont {Kohmoto}},\
  }\href {\doibase 10.1103/PhysRevB.34.2041} {\bibfield  {journal} {\bibinfo
  {journal} {Phys. Rev. B}\ }\textbf {\bibinfo {volume} {34}},\ \bibinfo
  {pages} {2041} (\bibinfo {year} {1986})}\BibitemShut {NoStop}%
\bibitem [{\citenamefont {Ostlund}\ \emph {et~al.}(1983)\citenamefont
  {Ostlund}, \citenamefont {Pandit}, \citenamefont {Rand}, \citenamefont
  {Schellnhuber},\ and\ \citenamefont
  {Siggia}}]{PhysRevLett.50.1873_Fibonacci_Model}%
  \BibitemOpen
  \bibfield  {author} {\bibinfo {author} {\bibfnamefont {S.}~\bibnamefont
  {Ostlund}}, \bibinfo {author} {\bibfnamefont {R.}~\bibnamefont {Pandit}},
  \bibinfo {author} {\bibfnamefont {D.}~\bibnamefont {Rand}}, \bibinfo {author}
  {\bibfnamefont {H.~J.}\ \bibnamefont {Schellnhuber}}, \ and\ \bibinfo
  {author} {\bibfnamefont {E.~D.}\ \bibnamefont {Siggia}},\ }\href {\doibase
  10.1103/PhysRevLett.50.1873} {\bibfield  {journal} {\bibinfo  {journal}
  {Phys. Rev. Lett.}\ }\textbf {\bibinfo {volume} {50}},\ \bibinfo {pages}
  {1873} (\bibinfo {year} {1983})}\BibitemShut {NoStop}%
\bibitem [{\citenamefont {Kohmoto}\ \emph {et~al.}(1983)\citenamefont
  {Kohmoto}, \citenamefont {Kadanoff},\ and\ \citenamefont
  {Tang}}]{PhysRevLett.50.1870_Fibonacci_Model}%
  \BibitemOpen
  \bibfield  {author} {\bibinfo {author} {\bibfnamefont {M.}~\bibnamefont
  {Kohmoto}}, \bibinfo {author} {\bibfnamefont {L.~P.}\ \bibnamefont
  {Kadanoff}}, \ and\ \bibinfo {author} {\bibfnamefont {C.}~\bibnamefont
  {Tang}},\ }\href {\doibase 10.1103/PhysRevLett.50.1870} {\bibfield  {journal}
  {\bibinfo  {journal} {Phys. Rev. Lett.}\ }\textbf {\bibinfo {volume} {50}},\
  \bibinfo {pages} {1870} (\bibinfo {year} {1983})}\BibitemShut {NoStop}%
\bibitem [{\citenamefont {Das~Sarma}\ \emph {et~al.}(1988)\citenamefont
  {Das~Sarma}, \citenamefont {He},\ and\ \citenamefont
  {Xie}}]{PhysRevLett.61.2144_modulated_harper_model}%
  \BibitemOpen
  \bibfield  {author} {\bibinfo {author} {\bibfnamefont {S.}~\bibnamefont
  {Das~Sarma}}, \bibinfo {author} {\bibfnamefont {S.}~\bibnamefont {He}}, \
  and\ \bibinfo {author} {\bibfnamefont {X.~C.}\ \bibnamefont {Xie}},\ }\href
  {\doibase 10.1103/PhysRevLett.61.2144} {\bibfield  {journal} {\bibinfo
  {journal} {Phys. Rev. Lett.}\ }\textbf {\bibinfo {volume} {61}},\ \bibinfo
  {pages} {2144} (\bibinfo {year} {1988})}\BibitemShut {NoStop}%
\bibitem [{\citenamefont {Szab\'o}\ and\ \citenamefont
  {Schneider}(2020)}]{PhysRevB.101.014205}%
  \BibitemOpen
  \bibfield  {author} {\bibinfo {author} {\bibfnamefont {A.}~\bibnamefont
  {Szab\'o}}\ and\ \bibinfo {author} {\bibfnamefont {U.}~\bibnamefont
  {Schneider}},\ }\href {\doibase 10.1103/PhysRevB.101.014205} {\bibfield
  {journal} {\bibinfo  {journal} {Phys. Rev. B}\ }\textbf {\bibinfo {volume}
  {101}},\ \bibinfo {pages} {014205} (\bibinfo {year} {2020})}\BibitemShut
  {NoStop}%
\bibitem [{\citenamefont {Bulaevskii}\ and\ \citenamefont
  {Sadovskii}(1985)}]{bulaevskii_anderson_1985}%
  \BibitemOpen
  \bibfield  {author} {\bibinfo {author} {\bibfnamefont {L.~N.}\ \bibnamefont
  {Bulaevskii}}\ and\ \bibinfo {author} {\bibfnamefont {M.~V.}\ \bibnamefont
  {Sadovskii}},\ }\href {\doibase 10.1007/BF00681506} {\bibfield  {journal}
  {\bibinfo  {journal} {Journal of Low Temperature Physics}\ }\textbf {\bibinfo
  {volume} {59}},\ \bibinfo {pages} {89} (\bibinfo {year} {1985})}\BibitemShut
  {NoStop}%
\bibitem [{\citenamefont {Sadovskii}(1997)}]{SADOVSKII1997225}%
  \BibitemOpen
  \bibfield  {author} {\bibinfo {author} {\bibfnamefont {M.~V.}\ \bibnamefont
  {Sadovskii}},\ }\href {\doibase
  https://doi.org/10.1016/S0370-1573(96)00036-1} {\bibfield  {journal}
  {\bibinfo  {journal} {Physics Reports}\ }\textbf {\bibinfo {volume} {282}},\
  \bibinfo {pages} {225 } (\bibinfo {year} {1997})}\BibitemShut {NoStop}%
\bibitem [{\citenamefont {Ketterson}\ and\ \citenamefont
  {Song}(1999)}]{KettersonBook1999}%
  \BibitemOpen
  \bibfield  {author} {\bibinfo {author} {\bibfnamefont {J.~B.}\ \bibnamefont
  {Ketterson}}\ and\ \bibinfo {author} {\bibfnamefont {S.~N.}\ \bibnamefont
  {Song}},\ }\href@noop {} {\emph {\bibinfo {title} {Superconductivity}}}\
  (\bibinfo  {publisher} {Cambridge University Press},\ \bibinfo {year}
  {1999})\BibitemShut {NoStop}%
\bibitem [{\citenamefont {Feigel'man}\ \emph {et~al.}(2010)\citenamefont
  {Feigel'man}, \citenamefont {Ioffe}, \citenamefont {Kravtsov},\ and\
  \citenamefont {Cuevas}}]{Fractal_superconductivity}%
  \BibitemOpen
  \bibfield  {author} {\bibinfo {author} {\bibfnamefont {M.}~\bibnamefont
  {Feigel'man}}, \bibinfo {author} {\bibfnamefont {L.}~\bibnamefont {Ioffe}},
  \bibinfo {author} {\bibfnamefont {V.}~\bibnamefont {Kravtsov}}, \ and\
  \bibinfo {author} {\bibfnamefont {E.}~\bibnamefont {Cuevas}},\ }\href
  {\doibase https://doi.org/10.1016/j.aop.2010.04.001} {\bibfield  {journal}
  {\bibinfo  {journal} {Annals of Physics}\ }\textbf {\bibinfo {volume}
  {325}},\ \bibinfo {pages} {1390 } (\bibinfo {year} {2010})},\ \bibinfo {note}
  {july 2010 Special Issue}\BibitemShut {NoStop}%
\bibitem [{\citenamefont {Feigel'man}\ \emph {et~al.}(2007)\citenamefont
  {Feigel'man}, \citenamefont {Ioffe}, \citenamefont {Kravtsov},\ and\
  \citenamefont {Yuzbashyan}}]{PhysRevLett.98.027001}%
  \BibitemOpen
  \bibfield  {author} {\bibinfo {author} {\bibfnamefont {M.~V.}\ \bibnamefont
  {Feigel'man}}, \bibinfo {author} {\bibfnamefont {L.~B.}\ \bibnamefont
  {Ioffe}}, \bibinfo {author} {\bibfnamefont {V.~E.}\ \bibnamefont {Kravtsov}},
  \ and\ \bibinfo {author} {\bibfnamefont {E.~A.}\ \bibnamefont {Yuzbashyan}},\
  }\href {\doibase 10.1103/PhysRevLett.98.027001} {\bibfield  {journal}
  {\bibinfo  {journal} {Phys. Rev. Lett.}\ }\textbf {\bibinfo {volume} {98}},\
  \bibinfo {pages} {027001} (\bibinfo {year} {2007})}\BibitemShut {NoStop}%
\bibitem [{\citenamefont {Burmistrov}\ \emph {et~al.}(2012)\citenamefont
  {Burmistrov}, \citenamefont {Gornyi},\ and\ \citenamefont
  {Mirlin}}]{PhysRevLett.108.017002}%
  \BibitemOpen
  \bibfield  {author} {\bibinfo {author} {\bibfnamefont {I.~S.}\ \bibnamefont
  {Burmistrov}}, \bibinfo {author} {\bibfnamefont {I.~V.}\ \bibnamefont
  {Gornyi}}, \ and\ \bibinfo {author} {\bibfnamefont {A.~D.}\ \bibnamefont
  {Mirlin}},\ }\href {\doibase 10.1103/PhysRevLett.108.017002} {\bibfield
  {journal} {\bibinfo  {journal} {Phys. Rev. Lett.}\ }\textbf {\bibinfo
  {volume} {108}},\ \bibinfo {pages} {017002} (\bibinfo {year}
  {2012})}\BibitemShut {NoStop}%
\bibitem [{\citenamefont {Mayoh}\ and\ \citenamefont
  {Garc\'{\i}a-Garc\'{\i}a}(2015)}]{PhysRevB.92.174526}%
  \BibitemOpen
  \bibfield  {author} {\bibinfo {author} {\bibfnamefont {J.}~\bibnamefont
  {Mayoh}}\ and\ \bibinfo {author} {\bibfnamefont {A.~M.}\ \bibnamefont
  {Garc\'{\i}a-Garc\'{\i}a}},\ }\href {\doibase 10.1103/PhysRevB.92.174526}
  {\bibfield  {journal} {\bibinfo  {journal} {Phys. Rev. B}\ }\textbf {\bibinfo
  {volume} {92}},\ \bibinfo {pages} {174526} (\bibinfo {year}
  {2015})}\BibitemShut {NoStop}%
\bibitem [{\citenamefont {Zhao}\ \emph {et~al.}(2019)\citenamefont {Zhao},
  \citenamefont {Lin}, \citenamefont {Xiao}, \citenamefont {Huang},
  \citenamefont {Yao}, \citenamefont {Yan}, \citenamefont {Xing}, \citenamefont
  {Zhang}, \citenamefont {Li}, \citenamefont {Hoshino}, \citenamefont {Wang},
  \citenamefont {Zhou}, \citenamefont {Gu}, \citenamefont {Bahramy},
  \citenamefont {Yao}, \citenamefont {Nagaosa}, \citenamefont {Xue},
  \citenamefont {Law}, \citenamefont {Chen},\ and\ \citenamefont
  {Ji}}]{zhao_disorder-induced_2019}%
  \BibitemOpen
  \bibfield  {author} {\bibinfo {author} {\bibfnamefont {K.}~\bibnamefont
  {Zhao}}, \bibinfo {author} {\bibfnamefont {H.}~\bibnamefont {Lin}}, \bibinfo
  {author} {\bibfnamefont {X.}~\bibnamefont {Xiao}}, \bibinfo {author}
  {\bibfnamefont {W.}~\bibnamefont {Huang}}, \bibinfo {author} {\bibfnamefont
  {W.}~\bibnamefont {Yao}}, \bibinfo {author} {\bibfnamefont {M.}~\bibnamefont
  {Yan}}, \bibinfo {author} {\bibfnamefont {Y.}~\bibnamefont {Xing}}, \bibinfo
  {author} {\bibfnamefont {Q.}~\bibnamefont {Zhang}}, \bibinfo {author}
  {\bibfnamefont {Z.-X.}\ \bibnamefont {Li}}, \bibinfo {author} {\bibfnamefont
  {S.}~\bibnamefont {Hoshino}}, \bibinfo {author} {\bibfnamefont
  {J.}~\bibnamefont {Wang}}, \bibinfo {author} {\bibfnamefont {S.}~\bibnamefont
  {Zhou}}, \bibinfo {author} {\bibfnamefont {L.}~\bibnamefont {Gu}}, \bibinfo
  {author} {\bibfnamefont {M.~S.}\ \bibnamefont {Bahramy}}, \bibinfo {author}
  {\bibfnamefont {H.}~\bibnamefont {Yao}}, \bibinfo {author} {\bibfnamefont
  {N.}~\bibnamefont {Nagaosa}}, \bibinfo {author} {\bibfnamefont {Q.-K.}\
  \bibnamefont {Xue}}, \bibinfo {author} {\bibfnamefont {K.~T.}\ \bibnamefont
  {Law}}, \bibinfo {author} {\bibfnamefont {X.}~\bibnamefont {Chen}}, \ and\
  \bibinfo {author} {\bibfnamefont {S.-H.}\ \bibnamefont {Ji}},\ }\href
  {\doibase 10.1038/s41567-019-0570-0} {\bibfield  {journal} {\bibinfo
  {journal} {Nature Physics}\ }\textbf {\bibinfo {volume} {15}},\ \bibinfo
  {pages} {904} (\bibinfo {year} {2019})},\ \bibinfo {note} {number: 9
  Publisher: Nature Publishing Group}\BibitemShut {NoStop}%
\bibitem [{\citenamefont {Ghosal}\ \emph {et~al.}(1998)\citenamefont {Ghosal},
  \citenamefont {Randeria},\ and\ \citenamefont
  {Trivedi}}]{PhysRevLett.81.3940}%
  \BibitemOpen
  \bibfield  {author} {\bibinfo {author} {\bibfnamefont {A.}~\bibnamefont
  {Ghosal}}, \bibinfo {author} {\bibfnamefont {M.}~\bibnamefont {Randeria}}, \
  and\ \bibinfo {author} {\bibfnamefont {N.}~\bibnamefont {Trivedi}},\ }\href
  {\doibase 10.1103/PhysRevLett.81.3940} {\bibfield  {journal} {\bibinfo
  {journal} {Phys. Rev. Lett.}\ }\textbf {\bibinfo {volume} {81}},\ \bibinfo
  {pages} {3940} (\bibinfo {year} {1998})}\BibitemShut {NoStop}%
\bibitem [{\citenamefont {Gastiasoro}\ and\ \citenamefont
  {Andersen}(2018)}]{PhysRevB.98.184510}%
  \BibitemOpen
  \bibfield  {author} {\bibinfo {author} {\bibfnamefont {M.~N.}\ \bibnamefont
  {Gastiasoro}}\ and\ \bibinfo {author} {\bibfnamefont {B.~M.}\ \bibnamefont
  {Andersen}},\ }\href {\doibase 10.1103/PhysRevB.98.184510} {\bibfield
  {journal} {\bibinfo  {journal} {Phys. Rev. B}\ }\textbf {\bibinfo {volume}
  {98}},\ \bibinfo {pages} {184510} (\bibinfo {year} {2018})}\BibitemShut
  {NoStop}%
\bibitem [{\citenamefont {Martin}\ \emph {et~al.}(2005)\citenamefont {Martin},
  \citenamefont {Podolsky},\ and\ \citenamefont
  {Kivelson}}]{PhysRevB.72.060502}%
  \BibitemOpen
  \bibfield  {author} {\bibinfo {author} {\bibfnamefont {I.}~\bibnamefont
  {Martin}}, \bibinfo {author} {\bibfnamefont {D.}~\bibnamefont {Podolsky}}, \
  and\ \bibinfo {author} {\bibfnamefont {S.~A.}\ \bibnamefont {Kivelson}},\
  }\href {\doibase 10.1103/PhysRevB.72.060502} {\bibfield  {journal} {\bibinfo
  {journal} {Phys. Rev. B}\ }\textbf {\bibinfo {volume} {72}},\ \bibinfo
  {pages} {060502} (\bibinfo {year} {2005})}\BibitemShut {NoStop}%
\end{thebibliography}%
\end{document}